\pdfoutput=1

\documentclass[11pt]{article}
\usepackage{epsf,amsmath,amssymb,graphicx,scalefnt,rotating,enumitem,fancyhdr}
\numberwithin{equation}{section}
\usepackage{acronym}

\usepackage[final]{showkeys}
\usepackage{cite}
\usepackage{filemod}
\usepackage{framed}
\usepackage{comment}
\usepackage[pdftex, colorlinks=true, linkcolor=black, filecolor=black,
  citecolor=black, urlcolor=black, draft=false, bookmarks,
  bookmarksnumbered=true, plainpages=false,
  linktocpage={true}]{hyperref}
\usepackage[affil-it]{authblk}
\usepackage[usenames,dvipsnames]{color}
\usepackage{slashed}
\usepackage{dsfont}

\newif\ifshownewacro
\includecomment{private}
\shownewacrotrue  %
\newcounter{notecount}

\newcommand{\logmuqms}{l_{\mu Q}}
\newcommand{\lmut}{L_{\mu t}}
\newcommand{\dilogv}{\text{Li}_2(1/4)}
\newcommand{\Zchiring}{\mathring{Z}_\chi}
\newcommand{\noeqn}[1]{(\ref{#1})}
\newcommand{\ccf}{C_\text{F}}

\newcommand{\ctr}{T_\text{R}}
\newcommand{\nc}{n_\text{c}}
\newcommand{\na}{n_\text{A}}
\newcommand{\nf}{n_\text{f}}
\newcommand{\nh}{n_\text{h}}
\newcommand{\nl}{n_\text{l}}
\newcommand{\cca}{C_\text{A}}

\newcommand{\ren}{\text{\abbrev{R}}}
\newcommand{\bare}{\text{\abbrev{B}}}
\newcommand{\citere}[1]{Ref.~\cite{#1}}
\newcommand{\citeres}[1]{Refs.~\cite{#1}}

\newcommand{\abbrev}[1]{{\scalefont{.9}#1}}
\newcommand{\EulerGamma}{\gamma_\text{E}}
\newcommand{\ep}{\epsilon}
\newcommand{\api}{a_s}
\newcommand{\apit}{\bar{a}_s}
\newcommand{\eqn}[1]{Eq.\,(\ref{#1})}
\newcommand{\eqs}[1]{Eqs.\,(\ref{#1})}
\newcommand{\fig}[1]{Fig.\,\ref{#1}}

\newcommand{\sct}[1]{Sect.\,\ref{#1}}
\newcommand{\app}[1]{Appendix\,\ref{#1}}

\newcommand{\dd}{\mathrm{d}}

\newcommand{\deriv}[3]{\frac{\partial\ifthenelse{\equal{#1}{}}{}{^{#1}}%
    #2}{\partial #3\ifthenelse{\equal{#1}{}}{}{^{#1}}}}
\newcommand{\dderiv}[3]{\frac{\dd\ifthenelse{\equal{#1}{}}{}{^{#1}}%
    #2}{\dd #3\ifthenelse{\equal{#1}{}}{}{^{#1}}}}
\newcommand{\order}[1]{\ensuremath{{\cal O}(#1)}}

\newcommand{\msbar}{\ensuremath{\overline{\mbox{\abbrev{MS}}}}}

\newcommand{\tcalo}{\tilde{\calo}}
\newcommand{\bcalo}{\bar{\calo}}
\newcommand{\calo}{O}

\newcommand{\myacrodef}[3]{\acrodef{#2}{#3}\newcommand{#1}{\ac{#2}}}
\newcommand{\hkl}{\abbrev{HKL}}
\acused{HKL} %
\myacrodef{\wrt}{w.r.t.}{with respect to}
\myacrodef{\vpf}{VPF}{vacuum polarization function}
\myacrodef{\vev}{VEV}{vacuum expectation value}
\myacrodef{\rg}{RG}{renormalization group}
\myacrodef{\gff}{GFF}{gradient-flow formalism}
\myacrodef{\ope}{OPE}{Operator Product Expansion}
\newcommand{\qcd}{\abbrev{QCD}}
\newcommand{\qed}{\abbrev{QED}}
\acused{QCD} %
\myacrodef{\lhc}{LHC}{Large Hadron Collider}
\myacrodef{\uv}{UV}{ultra-violet}
\myacrodef{\lo}{LO}{leading order}
\myacrodef{\nlo}{NLO}{next-to-leading order}
\myacrodef{\nnlo}{NNLO}{next-to-next-to-leading order}
\myacrodef{\llog}{LL}{leading logarithmic}
\myacrodef{\nll}{NLL}{next-to-leading logarithmic}
\myacrodef{\nnll}{NNLL}{next-to-next-to-leading logarithmic}
\myacrodef{\pdf}{PDF}{parton density function}
\myacrodef{\sm}{SM}{Standard Model}
\myacrodef{\bsm}{BSM}{beyond-the-\ac{SM}}
\myacrodef{\mssm}{MSSM}{Minimal Supersymmetric \ac{SM}}
\myacrodef{\susy}{SUSY}{Supersymmetry}
\myacrodef{\dreg}{DREG}{Dimensional Regularization}
\myacrodef{\dred}{DRED}{Dimensional Reduction}
\myacrodef{\emt}{EMT}{energy-momentum tensor}
\allowdisplaybreaks
\newcommand{\RHheaderline}{\textsf{FERMILAB-PUB-20-249-T,
    IIT-CAPP-20-01, TTK-20-20 --- July 2020}
}
\fancypagestyle{firstpage}
{
  
  \fancyhead[R]{\RHheaderline}
}
\title{Hadronic vacuum polarization using gradient flow}
\author[1]{Robert
  V. Harlander}
\author[1]{Fabian Lange}
\author[2,3]{Tobias Neumann}
\affil[1]{TTK, RWTH Aachen University, 52056 Aachen, Germany}
\affil[2]{Department of Physics, Illinois Institute of Technology,
  Chicago, Illinois 60616, USA}
\affil[3]{Fermilab, PO Box 500, Batavia, Illinois 60510, USA}

\date{}
\begin{document}
\maketitle
\thispagestyle{firstpage}
\begin{abstract}
  The gradient-flow operator product expansion for \qcd\ current
  correlators including operators up to mass dimension four is
  calculated through \abbrev{NNLO}.  This paves an alternative way for
  efficient lattice evaluations of hadronic vacuum polarization
  functions.  In addition, flow-time evolution equations for flowed
  composite operators are derived. Their explicit form for the
  non-trivial dimension-four operators of \qcd\ is given through order
  $\alpha_s^3$.
\end{abstract}
\parskip.0cm
\tableofcontents
\parskip.2cm

\section{Introduction}\label{sec:intro}

The \acp{VPF} for (axial-)vector and (pseudo-)scalar particles are among
the most important objects when studying \qcd. On the one hand, this is
because their imaginary part is directly related to physical observables
such as the decay rates of the $Z$- or the Higgs boson, or the hadronic
R-ratio. If the characteristic energy scale is far above the \qcd\
scale, a perturbative evaluation of the polarization functions is
sufficient in these cases to arrive at high-precision results~(see,
e.g., \citere{Chetyrkin:1994js}).

But \vpf{}s also contribute indirectly to physical observables such as
anomalous magnetic moments\cite{Jegerlehner:2017gek,Aoyama:2020ynm}, the
definition of short-distance quark masses\cite{Chetyrkin:2010ic}, or
hadronic contributions to the \qed\ coupling
\cite{crivellin:2020,Keshavarzi:2020bfy}.  These applications involve an
integration of the \vpf{}s over the non-perturbative regime, which is
typically achieved with the help of experimental data and dispersion
relations.  Only very recently, first-principle lattice calculations
have become competitive with these dispersive approaches. In the case of
the hadronic vacuum polarization contribution to the muon's anomalous
magnetic moment, the two approaches turn out to lead to incompatible
results\cite{Borsanyi:2020mff}\footnote{The lattice calculation of the
  light-by-light contribution to $(g-2)_\mu$ is in agreement with other
  determinations though\cite{Blum:2019ugy,Chao:2020kwq}.}. It would
therefore be highly desirable to have additional independent
first-principle calculations of the \vpf.

About ten years ago, the \gff\ was suggested as a mechanism to improve
the efficiency of lattice
calculations\cite{Luscher:2010iy,Luscher:2011bx,Luscher:2013cpa} (see
also \citeres{Narayanan:2006rf,Luscher:2009eq}).  Since then, it has
become a standard for the scale-setting
procedure\cite{Borsanyi:2012zs,Sommer:2014mea}. However, also other
applications of the \gff\ have been studied, among them a new way to
determine the energy-momentum tensor on the lattice. The underlying idea
in this case is the small-flow-time expansion of composite
operators\cite{Luscher:2011bx}, leading to the flowed \ope\ (also named
smeared \ope\ in \citere{Monahan:2014tea,Monahan:2015lha}), where the
regular operators are replaced by operators taken at finite flow
time. Its main advantages \wrt\ the regular \ope\ is the absence of
operator mixing, and the improved efficiency of the evaluation of
operator matrix elements. The translation of the regular to the flowed
operators can be done perturbatively. For the energy-momentum tensor, it
is available through
\nnlo{}\cite{Suzuki:2013gza,Makino:2014taa,Harlander:2018zpi}. Quite
recently, the small-flow-time expansion was applied at \nlo\ to
\abbrev{CP}-violating operators\cite{Rizik:2020naq}, and to four-quark
operators\cite{Suzuki:2020zue}.

In this paper, we present the flowed \ope\ for the time-ordered product
of two currents through \nnlo\ \qcd. Taking the \vev\
leads to the \vpf.  This should thus allow for an alternative
first-principle evaluation of \vpf{}s on the lattice. In addition, we
derive a general logarithmic flow-time evolution equation for flowed
operators which resembles the \rg\ equation of regular operators.

The remainder of this paper is organized as
follows. Section\,\ref{sec:ccc} introduces the regular \ope\ of current
correlators with operators up to mass dimension four. This includes the
renormalization of these operators as well as an overview of the
literature which provides the corresponding perturbative Wilson
coefficients. (The coefficients for the dimension-four operators for
various currents are reproduced in \app{app:coefs}.) The transition to
the flowed \ope\ is presented in \sct{sec:flope}. Section\,\ref{sec:mix}
describes the calculation of the mixing matrix between regular and
flowed operators in the small-flow-time limit. While a large part of
this mixing matrix is already known \cite{Harlander:2018zpi} and
recollected in \app{app:zeta22}, the missing components require higher
order mass terms of the \vev\ of the flowed dimension-four operators and
their renormalization with the help of the vacuum-energy renormalization
constant. These results complete the ingredients required for the flowed
\ope\ of the \vpf\ through \nnlo. In \sct{sec:evol}, we derive a
logarithmic evolution equation from the generic flow-time dependence of
the mixing matrix. Section\,\ref{sec:conclusions} presents our
conclusions and gives a short outlook on possible extensions of this
work.

\section{Current-current correlators}\label{sec:ccc}

Our results are presented for a general non-Abelian gauge theory based
on a simple compact Lie group with $\nf$ quark fields
$\psi_1,\ldots,\psi_{\nf}$ in the fundamental representation, of which
the first $\nh$ are degenerate with mass $m$, while the remaining $\nl$
are massless. The generators $T^a$ of the fundamental representation are
normalized as $\text{Tr}(T^aT^b)=-\ctr\delta^{ab}$, and the structure
constants $f^{abc}$ are defined through the Lie algebra $[T^a,T^b] =
f^{abc}T^c$.  The dimensions of the fundamental and the adjoint
representation are $\nc$ and $\na$, respectively, and their quadratic
Casimir eigenvalues are denoted by $\ccf$ and $\cca$.  For SU$(N)$, it
is
\begin{equation}\label{eq:58z02}
  \begin{split}
    \nc=N\,,\qquad \na=N^2-1\,,\qquad \ccf=\ctr\frac{N^2-1}{N}\,,\qquad
    \cca=2\ctr N\,,
  \end{split}
\end{equation}
and \qcd\ is recovered for $\ctr=1/2$ and $N=3$, i.e.\ $\ccf=4/3$ and
$\cca=3$.  For brevity, we often use ``\qcd'' also to refer to the more
general gauge theory in the following.

\subsection{Operator product expansion}\label{sec:ope}

\begin{figure}
 \begin{center}
   \begin{tabular}{c}
       \begin{tabular}{cc}
     \raisebox{0em}{%
       \includegraphics[%
         viewport = 160 590 405 730,clip,%
         height=.1\textheight]%
                       {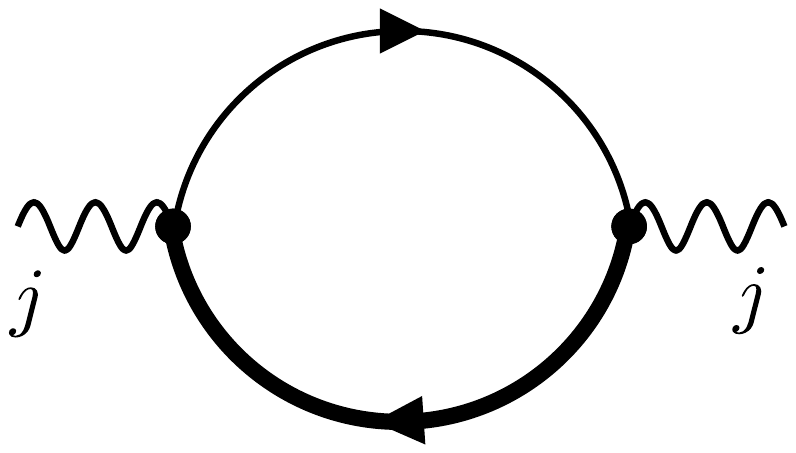}} &
     \raisebox{0em}{%
       \includegraphics[%
         viewport = 160 590 405 730,clip,%
         height=.1\textheight]%
                       {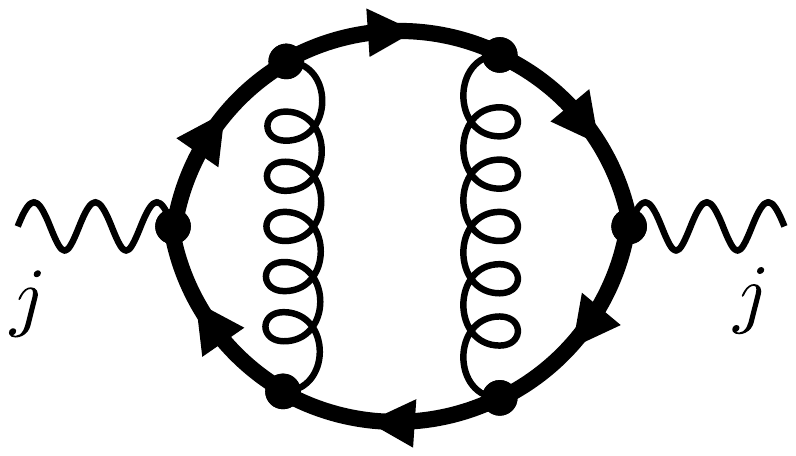}}\\
     (a) & (b)
       \end{tabular}
       \\
       \begin{tabular}{ccc}
     \raisebox{0em}{%
       \includegraphics[%
         viewport = 160 590 405 730,clip,%
         height=.1\textheight]%
                       {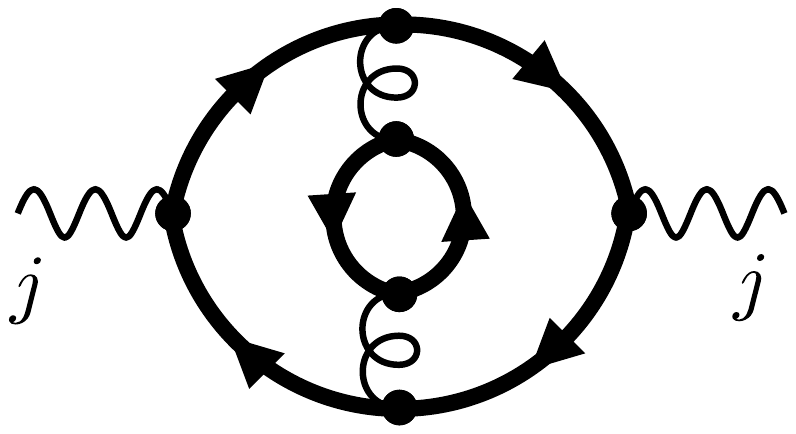}} &
     \raisebox{0em}{%
       \includegraphics[%
         viewport = 160 590 405 730,clip,%
         height=.1\textheight]%
                       {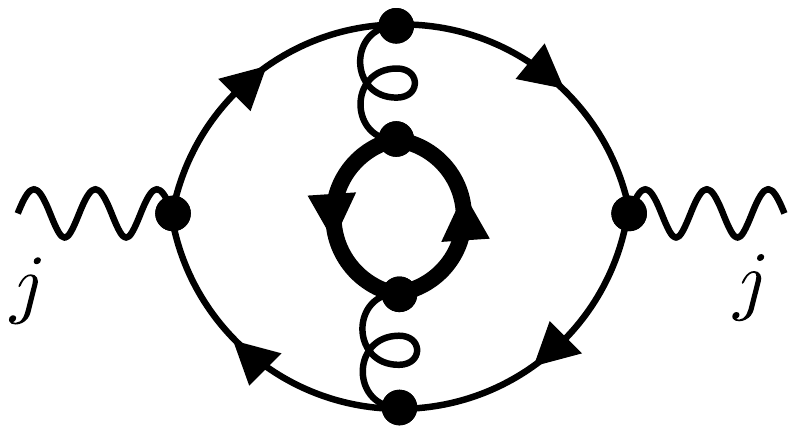}} &
     \raisebox{0em}{%
       \includegraphics[%
         viewport = 160 590 405 730,clip,%
         height=.1\textheight]%
                       {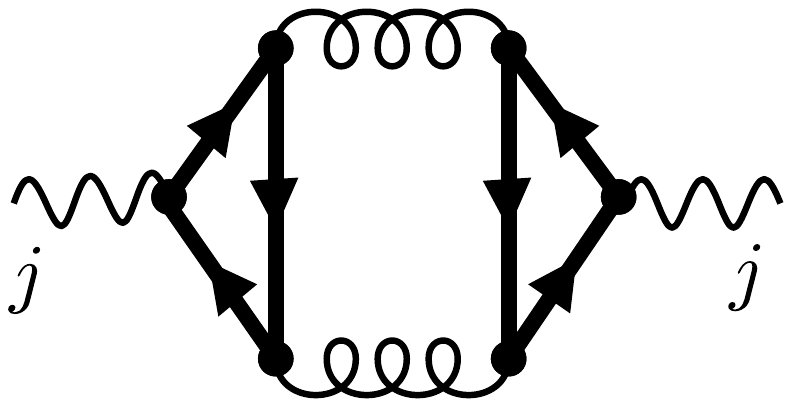}}
     \\
     (c) & (d) & (e)
       \end{tabular}
   \end{tabular}
   \parbox{.9\textwidth}{
     \caption[]{\label{fig:jj}\sloppy Sample diagrams contributing to
       the perturbative calculation of the \vpf, i.e., the left-hand
       side of \eqn{eq:ccope}. The currents are symbolized by wavy
       lines, gluons by spirals. We consider the case where $\nl$ quarks
       are massless (thin straight lines), and $\nh$ quarks are
       degenerate with mass $m$ (thick straight lines). (a)~One-loop
       contribution for non-diagonal currents; (b-e)~sample three-loop
       diagrams for diagonal currents. In (d), the currents couple to
       massless quarks. (e)~is a ``singlet'' diagram.  All diagrams in
       this paper were produced with the help of
       \texttt{FeynGame}\cite{Harlander:2020cyh}.}}
 \end{center}
\end{figure}

The role of the perturbative and the non-perturbative regime of
\acp{VPF} can be made most explicit through the \ope\ (see, e.g.,
\citere{Dominguez:2014vca}):
\begin{equation}
  \begin{split}
    T(Q) \equiv \int\dd^4 x\, e^{iQx} \langle Tj(x)j(0)\rangle
    \stackrel{Q^2\to \infty}{\sim} \sum_{k,n}
    C^{(k),\bare}_{n}(Q)\langle\calo^{(k)}_{n}(x=0)\rangle\,,
    \label{eq:ccope}
  \end{split}
\end{equation}
where $j(x)$ generically stands for a scalar, pseudo-scalar, vector,
axial-vector, or tensor current, and $k$ labels the mass
dimension. \fig{fig:jj} shows sample Feynman diagrams which arise from
the perturbative evaluation of the current correlator
in \eqn{eq:ccope}. In the following, we only consider the so-called
non-singlet diagrams, where the currents are connected by a common quark
line. An example for a singlet-diagram, on the other hand, is shown
in \fig{fig:jj}\,(e).

The coefficients $C_{n}^{(k),\bare}$ on the right-hand side of
\eqn{eq:ccope} depend on the quantum numbers of the current and may thus
carry Lorentz indices. Apart from the explicit results for specific
currents in \app{app:coefs}, we suppress these indices throughout the
paper.  We furthermore assume that, upon transition from the left- to
the right-hand side, possible global divergences are subtracted off of
$T(Q)$.

Up to mass dimension two, the only operators of \qcd\ which contribute
to physical matrix elements are proportional to unity, i.e.,
\begin{equation}\label{eq:dim02}
  \begin{split}
    \calo^{(0)}_{1} \equiv \calo^{(0)} = \mathds{1}\,,\qquad
    \calo^{(2)}_{1} \equiv \calo^{(2)} = m_\bare^2\,\mathds{1}\,,
  \end{split}
\end{equation}
where $m_\bare$ is the bare mass of the $\nh$ degenerate massive quarks.
This means that
\begin{equation}
  \begin{split}
    C_1^{(0)}\equiv
    C^{(0)}\equiv
    C^{(0),\bare}\,, \qquad
    C_1^{(2)}\equiv
    C^{(2)}\equiv
    Z_m^2C^{(2),\bare}
    \label{eq:dim2coefs}
  \end{split}
\end{equation}
are \uv-finite, where $Z_m$ is the renormalization constant of the quark
mass defined in \app{app:rg}.

At mass dimension four, we choose the following basis of operators (the
space-time argument is suppressed in most of what follows):
\begin{equation}
  \begin{split}
    \calo^{(4)}_{1} \equiv \calo_{1} &= \frac{1}{g_\bare^2}F^a_{\mu\nu}
    F^a_{\mu\nu}\,,\\
    \calo^{(4)}_{2} \equiv \calo_{2} &=
    \sum_{q=1}^{\nf}\bar\psi_q\overleftrightarrow{\slashed{D}}\psi_q\,,\\
    \calo^{(4)}_{3}\equiv \calo_{3} &=
    m_\bare^4\,,
    \label{eq:calo}
  \end{split}
\end{equation}
where
\begin{equation}\label{eq:overF}
  \begin{split}
     F^a_{\mu\nu} &= \partial_\mu A^a_\nu - \partial_\nu A^a_\mu +
     f^{abc}A_\mu^bA_\nu^c\,, \qquad\overleftrightarrow{D}\!_\mu =
     \partial_\mu - \overleftarrow{\partial}_\mu + 2A^a_\mu T^a\,,
  \end{split}
\end{equation}
with the regular (as opposed to ``flowed'', see \sct{sec:flope}) quark
and gluon fields $\psi_q(x)$ and $A^a_\mu(x)$, respectively, and the bare coupling
$g_\bare$.

We employ Euclidean space-time, but translation of the
intermediate formulas and the final results to Minkowski space is
possible without difficulty.  Working in
\begin{equation}\label{eq:p1gtg}
  \begin{split}
    d=4-2\ep
  \end{split}
\end{equation}
space-time dimensions, the mass dimensions of $\calo_1$ and $\calo_2$ are
actually equal to $d$, while that of $\calo_3$ is equal to 4. Higher
dimensional operators are neglected in the following.

The set in \eqn{eq:calo} does not contain gauge dependent operators, or
operators that vanish due to the equations of motion when acting on
physical states, since they are irrelevant for the scope of this
paper. In fact, in this respect the upper limit of the sum over $q$ in
$\calo_2$ could be replaced by $\nh$, because the terms with massless
quarks vanish on-shell. For the same reason, one could use $\calo_2'
\equiv -2 m_\bare\sum_{q=1}^{\nh}\bar\psi_q\psi_q$ instead of $\calo_2$
in the definition of the operator basis \noeqn{eq:calo}. Other choices
are possible as well, but the basis in \eqn{eq:calo} is particularly
suitable for our purposes, because it is most directly related to the
operator basis used in
\citeres{Makino:2014taa,Makino:2018rys,Harlander:2018zpi}.

Matrix elements of the dimension-four operators are divergent in
general. However, one may define ``renormalized operators''
$\calo_{n}^\ren$ as linear combinations among them, for which physical
matrix elements become finite:
\begin{equation}
  \begin{split}
    \calo^\ren_{n} = \sum_k Z_{nk}\,\calo_{k}\,.
    \label{eq:calob}
  \end{split}
\end{equation}
Analogously, one defines renormalized coefficient functions through the
condition
\begin{equation}\label{eq:ribfo}
  \begin{split}
    \sum_n C_n^\bare\calo_n \stackrel{!}{=}
    \sum_n C_n\calo^\ren_n\quad
    \Rightarrow\quad
    C_n = \sum_{m}C^\bare_m (Z^{-1})_{mn}\,,
  \end{split}
\end{equation}
where $C^\bare_n\equiv C_n^{(4),\bare}$, cf.\ \eqs{eq:ccope} and
\noeqn{eq:calo}.  It is well known that, since the operators of
\eqn{eq:calo} are part of the \qcd\ Lagrangian, the renormalization
matrix $Z$ can be expressed in terms of the anomalous dimensions of
\qcd\cite{Spiridonov:1984br,Spiridonov:1988md}:
\begin{equation}
  \begin{split}
    Z &= \left(
    \begin{matrix}
      Z_{2\times 2} & \vec{Z}_3\\[.5em]
      \vec{0}^{\,T} & Z_m^{-4}
    \end{matrix}
    \right)\,,\quad
\mbox{where}\quad
    Z_{2\times 2} = \left(
    \begin{matrix}
      -\ep/\beta_\ep & -2\gamma_m/\beta_\ep\\
      0 & 1
    \end{matrix}
    \right)\,,\\
      \vec{Z}_3 &=
      4\hat{\mu}^{-2\ep}Z_m^{-4}\left(
      \begin{matrix}
      \displaystyle \api\deriv{}{}{\api}Z_0 \\[1em]
      2\,Z_0
      \end{matrix}
      \right)\,,\quad \hat\mu\equiv
      \frac{\mu}{\sqrt{4\pi}}e^{\EulerGamma/2}\,.
      \label{eq:z}
  \end{split}
\end{equation}
The 't\,Hooft mass $\mu$ ensures that each renormalized operator
$\calo_n^\ren$ in \eqn{eq:calob} has the same mass dimension as the
corresponding bare one, and $\hat\mu$ appears because we will adopt the
\msbar\ scheme by default ($\EulerGamma=
-\Gamma'(1)=0.577216\ldots$). We also introduced the quantity
$\api=\alpha_s/\pi=g^2/(4\pi^2)$ here, where $g$ is the renormalized
strong coupling in the \msbar\ scheme.
$Z_0$ is the \msbar\ renormalization constant for the vacuum
energy\cite{Spiridonov:1988md}. It is given in \app{app:rg}, together
with the anomalous quark mass dimension $\gamma_m$ and the $d$-dimensional
beta function $\beta_\ep$.

\subsection{Coefficient functions}

The \ope\ form of the current correlators is usually inconvenient for
their perturbative evaluation. Instead, one rather evaluates the
l.h.s.\ of \eqn{eq:ccope} directly by calculating the relevant two-point
functions to the appropriate order. The exact analytical result for
general quark masses is known at the two-loop
level\cite{Kallen:1955fb,Broadhurst:1981jk,Kniehl:1989yc,Djouadi:1993ss},
while higher orders up to the four-loop level have been reconstructed by
combining various kinematical
limits\cite{Broadhurst:1993mw,Fleischer:1994ef,Broadhurst:1994qj,
  Baikov:1995ui,Chetyrkin:1995ii,Chetyrkin:1997mb,Chetyrkin:1998ix,
  Hoang:2008qy,Kiyo:2009gb,Greynat:2010kx,Greynat:2011zp}, or through
integral reduction\cite{Tkachov:1981wb,Chetyrkin:1981qh,Laporta:2001dd}
and subsequent numerical evaluation of the resulting master
integrals\cite{Maier:2017ypu}.

Since the dimension-zero and -two operators in \eqn{eq:ccope} are
proportional to unity, the coefficients $C^{(0)}$ and $C^{(2)}$ are
immediately determined from the small-mass expansion of these
perturbative results for the \vpf. They are thus known up to the
four-loop level at the
moment\cite{Chetyrkin:1979bj,Dine:1979qh,Celmaster:1979dm,
  Gorishnii:1986pz,Chetyrkin:1997qi,
  Harlander:1997kw,Harlander:1997xa,
  Baikov:2004ku,Baikov:2009uw}.\footnote{The imaginary parts of the
  \vpf{}s are known even at the five-loop level in the
  phenomenologically most relevant
  cases\cite{Chetyrkin:2005kn,Baikov:2005rw,Baikov:2012er,Baikov:2012zn}.}

The Wilson coefficients $C_n$ of the dimension-four operators, on the
other hand, require a dedicated calculation which keeps track of the
contributions from the individual operators. This has been done through
\order{\api^3} for $C_1$ and $C_2$, and through \order{\api^2} for $C_3$
in \citeres{Chetyrkin:1985kn,Chetyrkin:1994ex,
  Chetyrkin:2000zk,Harlander:diss}. For the purpose of this paper, only
the \order{\api^2} results are required. For completeness, we include
them in \app{app:coefs}.

\section{Flowed operator product expansion}\label{sec:flope}

Having introduced the setup in the ``regular'' theory, we now translate
this to the flowed \ope\ for the current correlators.

\subsection{Flowed operators}

We introduce the flowed operators as
\begin{equation}
  \begin{split}
    \tcalo_{1}(t,x) &=
    \frac{Z_s}{g_\bare^2}\,G^a_{\mu\nu}(t,x)G^a_{\mu\nu}(t,x) =
    \frac{\hat{\mu}^{-2\ep}}{g^2}\,G^a_{\mu\nu}(t,x)G^a_{\mu\nu}(t,x)\,,\\
    \tcalo_{2}(t,x)
    &= \Zchiring\sum_{q=1}^{\nf}\bar\chi_q(t,x)
    \overleftrightarrow{\slashed{\mathcal{D}}}(t,x)\chi_q(t,x)\,,\\[.5em]
    \tcalo_{3}(t,x) &= m^4\,,
    \label{eq:flopo}
  \end{split}
\end{equation}
where
\begin{equation}\label{eq:dleftright}
  \begin{split}
    \overleftrightarrow{\mathcal{D}}\!_\mu =
    {\mathcal{D}}\!_\mu -
    \overleftarrow{\mathcal{D}}\!_\mu\,,\qquad
                  {\mathcal{D}}\!_\mu = \partial_\mu + B^a_\mu T^a\,,\qquad
    \overleftarrow{\mathcal{D}}\!_\mu = 
                  \overleftarrow{\partial}_\mu - B^a_\mu T^a\,.
  \end{split}
\end{equation}
The flowed gauge and quark fields $B^a_\mu=B^a_\mu(x,t)$ and
$\chi_q=\chi_q(x,t)$ obey the equations\cite{Luscher:2010iy,Luscher:2013cpa}
\begin{equation}
  \begin{split}
    \partial_t B^a_\mu &= \mathcal{D}^{ab}_\nu G^b_{\nu\mu} + \kappa
    \mathcal{D}^{ab}_\mu \partial_\nu B^b_\nu\,,\\ \partial_t \chi_q
    &= \Delta \chi_q - \kappa \partial_\mu B^a_\mu T^a \chi_q\,,\\ \partial_t
    \bar \chi_q &= \bar \chi_q \overleftarrow \Delta + \kappa \bar
    \chi_q \partial_\mu B^a_\mu T^a\,,
    \label{eq:flow}
  \end{split}
\end{equation}
with the initial conditions
\begin{equation}
  \begin{split}
    B^a_\mu (t=0,x) = A^a_\mu (x)\,,\qquad \chi_q (t=0,x)= \psi_q(x)\,,\quad
    q\in\{1,\ldots \nf\}\,.
    \label{eq:bound}
  \end{split}
\end{equation}
Here we used the flowed covariant derivative in the adjoint
representation,
\begin{equation}\label{eq:dp3bb}
  \begin{split}
  \mathcal{D}_\mu^{ab} = \delta^{ab} \partial_\mu - f^{abc} B_\mu^c\,,
  \end{split}
\end{equation}
and the flowed Laplace operators 
\begin{equation}
  \begin{split}
    \Delta = \mathcal{D}_\mu \mathcal{D}_\mu
    \,,\qquad
    \overleftarrow{\Delta} = \overleftarrow{\mathcal{D}}_\mu
    \overleftarrow{\mathcal{D}}_\mu \,,
  \end{split}
\end{equation}
where the flowed covariant derivatives in the fundamental representation
are given in \eqn{eq:dleftright}. For our three-loop calculations, we
set the gauge parameter $\kappa=1$ in \eqn{eq:flow}, because this
simplifies the intermediate expressions. All our results are independent
of this choice though.

$\Zchiring$ is the non-minimal renormalization constant for the flowed
quark fields $\chi_q$, defined by the all-order
condition\cite{Makino:2014taa}
\begin{equation}\label{eq:zchidef}
  \begin{split}
    \langle \tcalo_2(t)\rangle\bigg|_{m=0} \equiv -\frac{2\nc\nf}{(4\pi t)^2}\,,
  \end{split}
\end{equation}
where $\langle\cdot\rangle$ denotes the \vev.  It reads
\begin{equation}\label{eq:608qo}
  \begin{split}
    \Zchiring = \zeta_\chi\,Z_\chi\,,
  \end{split}
\end{equation}
where $Z_\chi$ is the \msbar\ part,
\begin{equation}
  \begin{split}
    Z_\chi &=
    1
      + \api\frac{\gamma_{\chi,0}}{2\ep}
      + \api^2\left[\frac{\gamma_{\chi,0}}{4\ep^2}\left(
        \frac{\gamma_{\chi,0}}{2} - \beta_0\right)
        +\frac{\gamma_{\chi,1}}{4\ep}\right] + \order{\api^3}\,,
    \label{eq:zchi}
  \end{split}
\end{equation}
and
\begin{equation}
  \begin{split}
\zeta_\chi& = 1 + \api
          \left( \frac{\gamma_{\chi, 0}}{2} \lmut - \frac{3}{4} \ccf \ln 3
          - \ccf \ln 2 \right) \\ &
                \qquad + \api^2 \Bigg\{
                  \frac{\gamma_{\chi, 0}}{4} \left( \beta_0 +
                  \frac{\gamma_{\chi, 0}}{2} \right) \lmut^2 +
                  \Big[ \frac{\gamma_{\chi, 1}}{2} -
                  \frac{\gamma_{\chi, 0}}{2} \left( \beta_0 +
                  \frac{\gamma_{\chi, 0}}{2} \right) \ln 3 \\
                & \qquad \qquad -
                  \frac{2}{3}
                  \gamma_{\chi, 0} \left(
                  \beta_0 + \frac{\gamma_{\chi, 0}}{2}\right) \ln 2
                  \Big] \lmut
                  + \frac{c_\chi^{(2)}}{16} \Bigg\} + \order{\api^3}\,.
                \label{eq:zfchi}
  \end{split}
\end{equation}
The short-hand notation
\begin{equation}\label{eq:lmut}
  \begin{split}
    \lmut = \ln\frac{\mu^2}{\mu_t^2}\,,\qquad
    \mu_t = \frac{1}{\sqrt{2te^{\EulerGamma}}}
  \end{split}
\end{equation}
reflects our choice of the ``central'' renormalization scale
$\mu_t$\,\cite{Harlander:2018zpi}.

The minimal renormalization constant $Z_\chi$ is known analytically
through \nnlo\cite{Luscher:2013cpa,Harlander:2018zpi},
\begin{equation}
  \begin{split}
  \gamma_{\chi,0} &= \frac{3}{2}\ccf\,,\\
  \gamma_{\chi, 1} &= \ccf\left[
  \cca \left( \frac{223}{48} - \ln 2 \right) - \ccf \left( \frac{3}{16} + 
  \ln 2 \right) - \frac{11}{12} \ctr\nf\right]\,.
  \label{eq:gammachi}
  \end{split}
\end{equation}
The finite coefficient in \eqn{eq:zfchi} has been obtained
numerically in \citere{Artz:2019bpr}:
\begin{equation}\label{eq:n26a3}
  \begin{split}
    c_\chi^{(2)} &= \cca\ccf\,c_{\chi,\text{A}} + \ccf^2\,c_{\chi,\text{F}}
    +\ccf\ctr\nf\,c_{\chi,\text{R}}\,,
  \end{split}
\end{equation}
with\footnote{The sign on the r.h.s.\ of equation (B.3) in
  \citere{Artz:2019bpr} is incorrect. }
\begin{equation}\label{eq:c2rep}
  \begin{split}
    c_{\chi,\text{A}} &= -23.7947\,,\qquad
    c_{\chi,\text{F}} = 30.3914 \,,\\
    c_{\chi,\text{R}} &= -\frac{131}{18} + \frac{46}{3}\,\zeta(2)
    +\frac{944}{9}\ln 2 + \frac{160}{3}\,\ln^22 - \frac{172}{3}\,\ln3 +
 \frac{104}{3}\,\ln2\ln3\\&\quad - \frac{178}{3}\,\ln^23+
 \frac{8}{3}\,\text{Li}_2(1/9)-
 \frac{400}{3}\,\text{Li}_2(1/3) + \frac{112}{3}\,\text{Li}_2(3/4)
 = -3.92255\ldots\,,
  \end{split}
\end{equation}
with Riemann's zeta function $\zeta(s)\equiv \sum_{n=1}^\infty n^{-s}$
and the di-logarithm $\text{Li}_2(z)=\sum_{k=1}^\infty z^k/k^2$.
The strong coupling renormalization constant $Z_s$ in
\eqn{eq:flopo} ensures that matrix elements of $\tcalo_{1}(t)$ are
finite \cite{Luscher:2010iy,Luscher:2011bx}. The reason for keeping
track of the non-integer mass dimension of $\tcalo_1(t)$ is clarified
later.

\eqn{eq:c2rep} displays only the first six leading digits in numerical
results. Results with higher accuracy are provided in an ancillary file,
which also includes the $\lmut$ terms, see \app{app:anc}. We expect that
these floating-point numbers can be considered equivalent to their exact
results for all practical purposes. This is why we often use the
numerical values for the coefficients in what follows, even if the exact
result is available.

Similar to the regular operators in \eqn{eq:calo}, one could trade the
flowed operator $\tcalo_2(t)$ for $\tcalo_2'(t) =
-2m\Zchiring\sum_{q=1}^{\nf}\bar\chi_q(t)\chi_q(t)$. However, in this case the
final results to be derived below would be different, because the
equations of motion for the flowed operators relate $\tcalo'_2(t)$ to
both $\tcalo_2(t)$ and $\tcalo_1(t)$ (see
\citeres{Makino:2014taa,Harlander:2018zpi}). A transformation of the
results in this paper to $\tcalo'_2(t)$ is straightforward though.

\subsection{Small-flow-time expansion}

The small-flow-time expansion allows us to relate the \qcd\ operators
and coefficients with the flowed quantities as follows:
\begin{equation}
  \begin{split}
    \tcalo_{n}(t) &= \zeta^{(0)}_{n}(t)\mathds{1} +
    \zeta^{(2),\bare}_{n}(t)\,m_\bare^2\mathds{1} +
    \sum_{k}\zeta^\bare_{nk}(t)\calo_{k} + \ldots\\
     &\equiv \zeta^{(0)}_{n}(t)\mathds{1} +
    \zeta^{(2)}_{n}(t)\,m^2\mathds{1} +
    \sum_{k}\zeta_{nk}(t)\calo^\ren_{k} + \ldots\,,
    \label{eq:smalltexp}
  \end{split}
\end{equation}
where the ellipsis denotes terms that vanish as $t\to 0$,
and
\begin{equation}\label{eq:zetax}
  \begin{split}
    \zeta^{(2)}_{n}(t) = \zeta^{(2),\bare}_{n}(t)\,Z_m^2\,,\qquad
    \zeta_{nk}(t) = \sum_{l}\zeta_{nl}^\bare(t)\,Z_{lk}^{-1}
  \end{split}
\end{equation}
are the renormalized, finite mixing coefficients.  Inversion of
\eqn{eq:smalltexp} gives
\begin{equation}\label{eq:opoinv}
  \begin{split}
    \calo^\ren_{n} &= \sum_k \zeta_{nk}^{-1}(t)\,\bcalo_{k}(t) + \ldots\,,\\
    \bcalo_{n}(t) &\equiv \tcalo_{n}(t) - \zeta^{(0)}_{n}(t)\mathds{1}
    - \zeta^{(2)}_{n}(t)m^2\mathds{1}\,,
  \end{split}
\end{equation}
where $\zeta^{-1}_{nk}$ is the $nk$-element of the inverse of the mixing
matrix $\zeta$. This lets one define the ``flowed \ope'' for the
current correlator:
\begin{equation}
  \begin{split}
    T(Q)\stackrel{Q^2\to\infty}{\sim}
     \tilde{C}^{(0)}(Q^2,t)\mathds{1}
    + \tilde{C}^{(2)}(Q^2,t)m^2\mathds{1}
    + \sum_n\tilde{C}_{n}(Q^2,t)\tcalo_{n}(t) + \ldots
    \label{eq:flowedope}
  \end{split}
\end{equation}
where the corresponding coefficient functions are related to the regular
Wilson coefficients through
\begin{equation}
  \begin{split}
  \tilde C_{n}(Q^2,t) &= \sum_k C_{k}(Q^2)\zeta_{kn}^{-1}(t)\,,
    \\
  \tilde C^{(0,2)}(Q^2,t) &= C^{(0,2)}(Q^2) -
  \sum_{n}\tilde C_{n}(t,Q^2)\,\zeta^{(0,2)}_n(t)\,.
    \label{eq:ctilde}
  \end{split}
\end{equation}
The regular \qcd\ coefficients $C^{(0)}$ and $C^{(2)}$ are given by the
first two terms in $m^2/Q^2$ of the large-$Q^2$ expansion of the
\vpf{}s. Through the required order, they can be found in
\citere{Chetyrkin:1997qi} for vector-, in \citere{Harlander:1997kw} for
axial-vector-, and in \citere{Harlander:1997xa} for scalar- and pseudo-scalar
currents, for example. The dimension-four coefficients can be found in
\citeres{Chetyrkin:1985kn,Harlander:diss}. For convenience of the
reader, they are also collected in \app{app:coefs}.

\section{Calculation of the mixing matrix}\label{sec:mix}

We now determine the mixing matrix $\zeta$ in a perturbative calculation
through \nnlo. By using the known results for the regular Wilson
coefficients given in \app{app:coefs}, one can determine the flowed
coefficients of \eqn{eq:ctilde} to the same order. Together with an
evaluation of the flowed operator matrix elements on the lattice,
the \vpf{}s can be extracted and used in the determination of various
physical quantities.

The bare mixing matrix $\zeta^\bare$ can be determined with the help of
the \textit{method of projectors}:
\begin{equation}
  \begin{split}
    \zeta^{(0,2),\bare}_{n}(t) &= P^{(0,2)}[\tcalo_{n}(t)]\,,
\qquad    \zeta^\bare_{nk}(t) = P^{(4)}_{k}[\tcalo_{n}(t)]\,,
    \label{eq:projectors}
  \end{split}
\end{equation}
where the action of $P^{(n)}$ is to take suitable derivatives of a
specific Green's function of the operator such that
\begin{equation}\label{eq:kr5sv}
  \begin{split}
    P^{(n)}[\calo^{(m)}]=\delta_{nm}\,,\qquad
    P^{(n)}[\calo_k]=P^{(4)}_k[\calo^{(n)}]=0\,,\qquad
    P^{(4)}_k[\calo_l]=\delta_{kl}\,,
  \end{split}
\end{equation}
for $n,m\in\{0,2\}$ and $k,l\in\{1,2,3\}$.  For details, see
\citeres{Gorishnii:1983su,Gorishnii:1986gn,Harlander:2018zpi}.

Specifically, the projectors onto $\mathds{1}$, $m_\bare^2\mathds{1}$,
and $\calo^\bare_{3}$ are given by derivatives of vacuum matrix elements
\wrt\ $m_\bare$:
\begin{equation}\label{eq:zeta02}
  \begin{split}
    \zeta^{(0)}_{n}(t) &= P^{(0)}[\tcalo_{n}(t)] \equiv
    \langle\tcalo_{n}(t)\rangle\Big|_{m_\bare=0}\,,\\ \zeta^{(2)}_{n}(t)
    &= Z_m^2\,P^{(2)}[\tcalo_{n}(t)] \equiv Z_m^2\frac{1}{2!}\deriv{2}{}{m_\bare}
    \langle\tcalo_{n}(t)\rangle\Big|_{m_\bare=0}\,,\\ \zeta^\bare_{n3}(t)
    &= P^{(4)}_{3}[\tcalo_{n}(t)]\equiv \frac{1}{4!}\deriv{4}{}{m_\bare}
    \langle\tcalo_{n}(t)\rangle\Big|_{m_\bare=0}\,.
  \end{split}
\end{equation}
A crucial point of the \textit{method of projectors} is that the
derivatives and limits are taken \textit{before} loop
integration\cite{Gorishnii:1983su,Gorishnii:1986gn}. This guarantees
that all loop contributions of projections on the r.h.s.\
of \eqn{eq:smalltexp} lead to scaleless integrals and thus vanish in
dimensional regularization.  On the other hand, it implies that the
projections on the l.h.s.\ of \eqn{eq:smalltexp} can be divergent, even
though physical matrix elements of $\tcalo_n(t)$ are finite. This is why
we need to carefully account for a possible non-integer mass dimension
of these operators, see \eqn{eq:flopo}.

We directly obtain
\begin{equation}\label{eq:zeros}
  \begin{split}
    \zeta^\bare_{33} = \frac{1}{4!}\deriv{4}{}{m_\bare}\,m^4 = Z_m^{-4}
    \,,\qquad
    \zeta^{(0),\bare}_{3} = \zeta^{(2),\bare}_{3} = 0\,,\qquad
    \zeta^\bare_{31} =
    \zeta^\bare_{32} = 0\,,
  \end{split}
\end{equation}
where the third set of equations follows from $\tcalo_{3}(t)=m^4=\calo_{3}$
and the projector property
$P^{(4)}_{1}[\calo_{3}]=P^{(4)}_{2}[\calo_{3}]=0$, see \eqn{eq:kr5sv}.

The bare and renormalized mixing matrices for the dimension-four operators
thus take the form
\begin{equation}
  \begin{split}
    \zeta^\bare = \left(\begin{matrix}
      \zeta^\bare_{2\times 2} &\vec{\zeta}_3^{\,\bare}\\[.5em]
      \vec{0}^{\,T} & Z_m^{-4}
    \end{matrix}\right)\,,\qquad
    \zeta = \left(\begin{matrix}
      \zeta_{2\times 2} &\vec{\zeta}_3\\[.5em]
      \vec{0}^{\,T} & 1
    \end{matrix}\right)\,,
    \label{eq:zetabare}
  \end{split}
\end{equation}
where $\vec{0}^{\,T} = (0,0)$, and
\begin{equation}
\begin{aligned}
  \zeta^\bare_{2\times 2} &= \left(\begin{matrix}
    \zeta^\bare_{11} & \zeta^\bare_{12}\\[.5em]
    \zeta^\bare_{21} & \zeta^\bare_{22}
  \end{matrix}\right)\,,\qquad&
  \zeta_{2\times 2} &= \zeta^\bare_{2\times 2}Z_{2\times 2}^{-1}\,,\\
  \vec{\zeta}_3^{\,\bare} &= 
  \left(
  \zeta_{13}^\bare, \zeta_{23}^\bare
  \right)^T\,,\qquad&
  \vec{\zeta}_3 &= (\vec{\zeta}^{\,\bare}_3-\zeta_{2\times 2}\vec{Z}_3) Z_m^4\,.
\end{aligned}
  \label{eq:zetacomp}
\end{equation}
$\zeta_{2\times 2}$ can be obtained from the mixing matrix of the
operators occurring in the energy-momentum tensor and is accordingly
known through \nnlo\ \cite{Harlander:2018zpi}. Explicit results are
given in \app{app:zeta22}.

The coefficient $\zeta^{(0)}_{n}$ is simply the \vev\ of $\tcalo_{n}(t)$
for $m=0$. For $\tcalo_1(t)$ it has been calculated through \nnlo\ in
\citeres{Luscher:2010iy,Harlander:2016vzb,Artz:2019bpr}:\footnote{The
  coefficient of the $\cca^2$ term in \citere{Artz:2019bpr} contains a
  typo: instead of $27.9786$, it should read $27.9784$.}
\begin{equation}\label{eq:zeta01}
  \begin{split}
    \zeta^{(0)}_1(t) = \langle\tcalo_1(t)\rangle\bigg|_{m=0} &=
      \frac{3}{4\pi^2 t^2}\frac{\na}{8}\bigg\{ 1
    +\apit\bigg[
\cca\,\bigg(
      \frac{13}{9} 
      + \frac{11}{6}\,\ln2 
      - \frac{3}{4}\,\ln3
      \bigg) 
      - \frac{2}{9}\,\ctr\nf
      \bigg]\\
    &\quad
    +\apit^2\bigg[
      1.74865\,\cca^2 - 1.97283\,\cca\ctr\nf\\&\quad\qquad+
      0.306224\,\ccf\ctr\nf + 0.121042\,\ctr^2\nf^2
      \bigg]\bigg\} +\order{\apit^3}\,,
\end{split}
\end{equation}
where $\apit = \api(\mu_t)$. Due to the
\rg\ invariance of
$\api\tcalo_1(t)$\cite{Luscher:2010iy,Luscher:2011bx}, the result for
general values of the 't\,Hooft mass $\mu$ can be obtained by
multiplying the result given in \eqn{eq:zeta01} by $\apit/\api$,
replacing
\begin{equation}\label{eq:cx3w6}
  \begin{split}
    \apit = \api\bigg[1 + \api\beta_0\lmut + \api^2\lmut\left(
      \beta_1 + \beta_0^2\lmut\right)\bigg] + \order{\api^3}\,,
  \end{split}
\end{equation}
and re-expanding in $\api$.

For $\tcalo_2$, on the other hand, we have \eqn{eq:zchidef} to all
orders in perturbation theory by definition, i.e.
\begin{equation}\label{eq:9s23o}
  \begin{split}
    \zeta_2^{(0)}(t)=\langle \tcalo_2(t)\rangle\bigg|_{m=0} \equiv
    -\frac{2\nc\nf}{(4\pi t)^2}\,.
  \end{split}
\end{equation}

\begin{figure}
  \begin{center}
    \begin{tabular}{cccc}
      (a) &
      (b) &
      (c) &
      (d) \\
      \raisebox{0em}{%
        \includegraphics[%
          viewport = 190 500 390 710,clip,%
          height=.15\textheight]%
                        {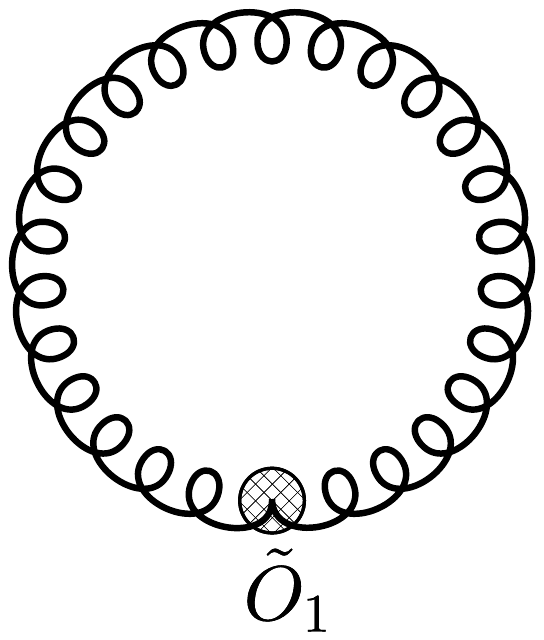}} &
      \raisebox{0em}{%
        \includegraphics[%
          viewport = 190 500 390 710,clip,%
          height=.15\textheight]%
                        {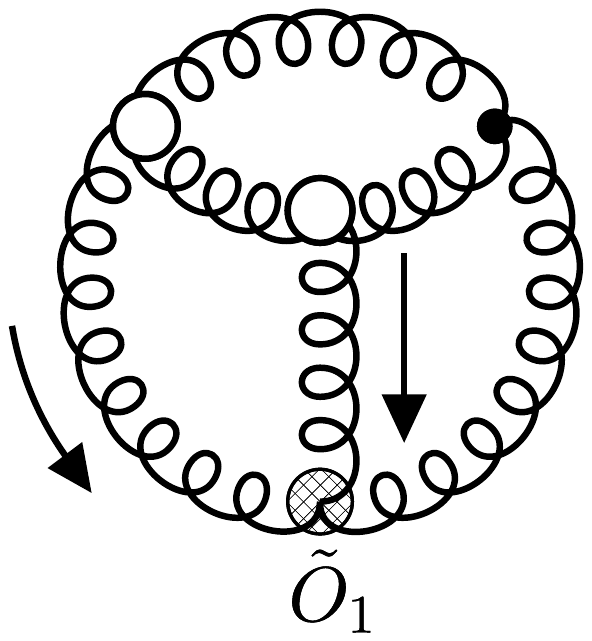}}&
      \raisebox{0em}{%
        \includegraphics[%
          viewport = 190 500 390 710,clip,%
          height=.15\textheight]%
                        {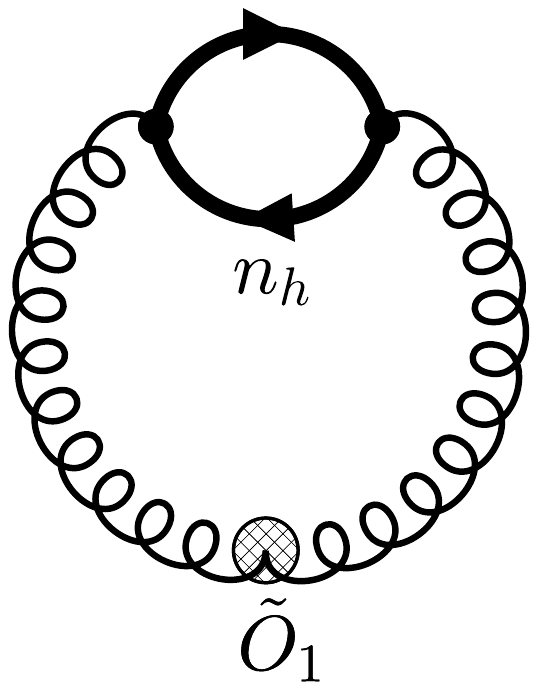}} &
      \raisebox{0em}{%
        \includegraphics[%
          viewport = 190 500 390 710,clip,%
          height=.15\textheight]%
                        {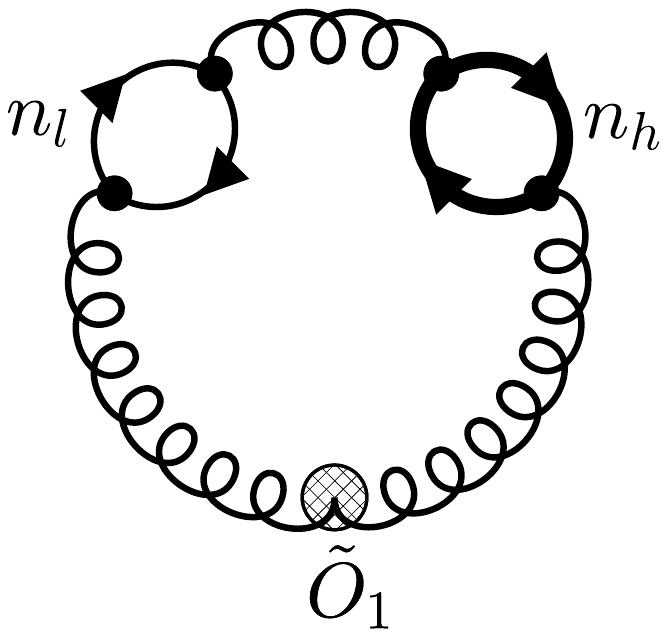}}\\[1.5em]
      (e) &
      (f) &
      (g) &
      (h) \\
      \raisebox{0em}{%
        \includegraphics[%
          viewport = 190 500 390 710,clip,%
          height=.15\textheight]%
                        {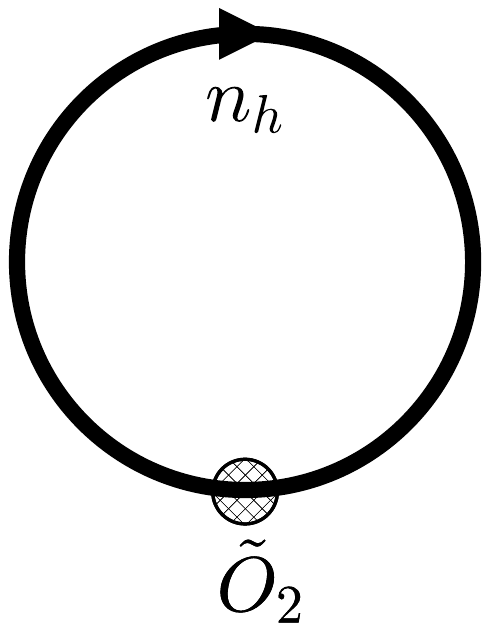}} &
      \raisebox{0em}{%
        \includegraphics[%
          viewport = 190 500 390 710,clip,%
          height=.15\textheight]%
                        {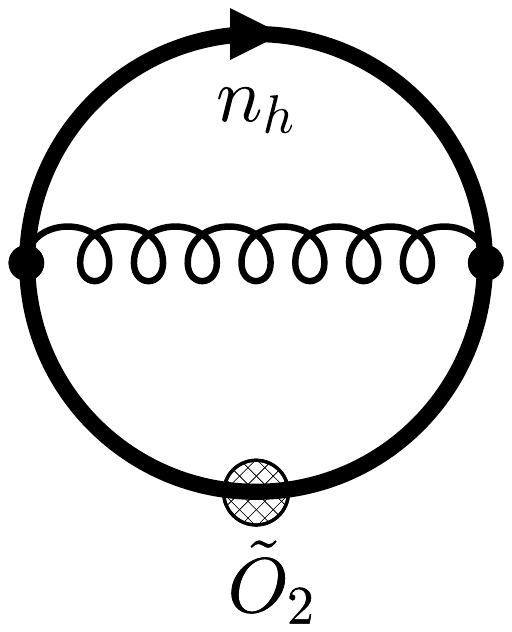}}&
      \raisebox{0em}{%
        \includegraphics[%
          viewport = 190 500 390 710,clip,%
          height=.15\textheight]%
                        {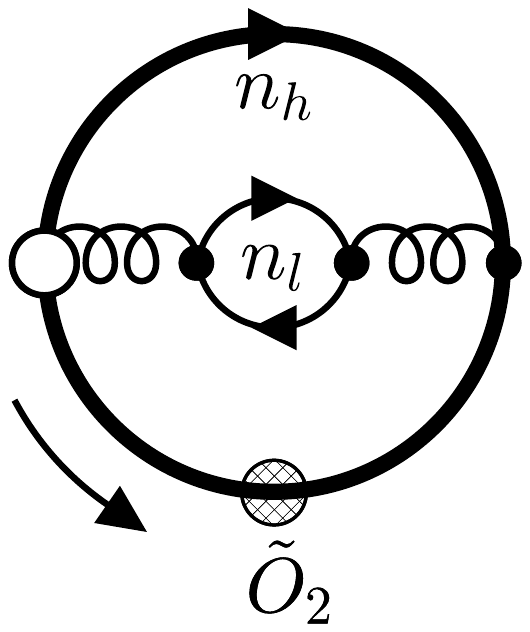}} &
      \raisebox{0em}{%
        \includegraphics[%
          viewport = 190 500 390 710,clip,%
          height=.15\textheight]%
                        {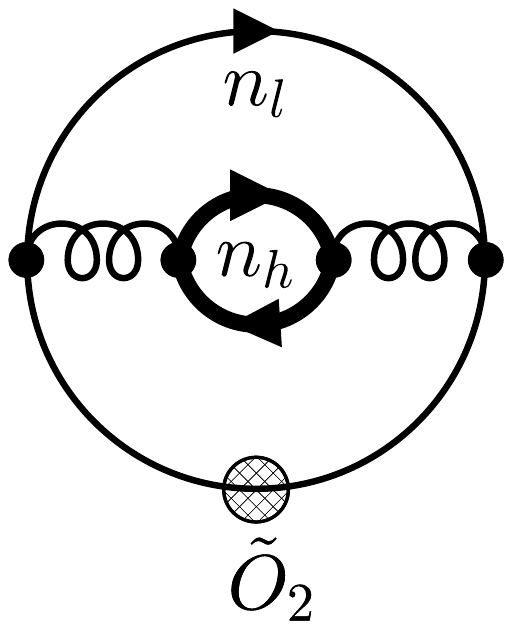}}
    \end{tabular}
    \parbox{.9\textwidth}{
      \caption[]{\label{fig:e1}\sloppy Sample diagrams contributing to
        $\zeta_{n}^{(m)}$ ($m\in\{0,2\}$) and $\zeta_{n3}$, for $n=1$
        (a-d) and $n=2$ (e-h). The notation is the same as in
        \fig{fig:jj}; in addition, white circles denote flow-vertices,
        and lines with arrows next to them denote flow-lines (see
        \citere{Artz:2019bpr} for details). Diagrams (a) and (b) only
        contribute to $\zeta_{1}^{(0)}$. }}
  \end{center}
\end{figure}

Therefore, the only coefficients that are not yet known through
\nnlo\ are
\begin{equation}\label{eq:unknownzetas}
  \begin{split}
\zeta^{(2)}_{1},\, \zeta^{(2)}_{2},\, \zeta_{13}\,\text{ and }
\zeta_{23}\,.
  \end{split}
\end{equation}
According to \eqn{eq:zeta02}, they require the calculation of
derivatives \wrt\ $m_\bare$ in $\langle\tcalo_1(t)\rangle$ and
$\langle\tcalo_2(t)\rangle$ up to three loops (see \fig{fig:e1} for a
set of associated diagrams). We achieved this using the setup described
in \citere{Artz:2019bpr}, which employs
\texttt{qgraf}\cite{Nogueira:1991ex,Nogueira:2006pq} and
\texttt{q2e/exp}\cite{Harlander:1997zb,Seidensticker:1999bb} for the
generation and subsequent categorization of the Feynman diagrams,
\texttt{FORM}\,\cite{Vermaseren:2000nd,Kuipers:2012rf} for the
manipulation of the ensuing algebraic expressions, the \texttt{color}
package\cite{vanRitbergen:1998pn} of \texttt{FORM} for the calculation
of the gauge group factors, and
\texttt{Kira}$\oplus$\texttt{FireFly}\cite{Maierhoefer:2017hyi,
  Maierhoefer:2018gpa,Klappert:2020nbg,Klappert:2019emp,Klappert:2020aqs} for
the Feynman integral reduction using integration-by-parts identities and
the Laporta
algorithm\cite{Tkachov:1981wb,Chetyrkin:1981qh,Laporta:2001dd} over
finite fields\cite{Kauers:2008zz,Kant:2013vta,vonManteuffel:2014ixa,Peraro:2016wsq}. For the evaluation of the master
integrals, we adopt the method described in \citere{Harlander:2016vzb},
which performs sector decomposition\,\cite{Binoth:2003ak} with the help
of \texttt{FIESTA}\,\cite{Smirnov:2013eza,Smirnov:2015mct} in order to
extract the \uv\ poles, along with a fully symmetric integration rule of
order 13 for the numerical evaluation of their
coefficients\cite{Genz:1983malik}, implemented with high precision
arithmetics by using the \texttt{MPFR} library\cite{mpfr:2007}.  Some
intermediate steps of the calculation are done within
\texttt{Mathematica}\cite{Mathematica}.

Multiplying the result by $Z_m^2$ suffices to obtain the renormalized
expressions for $\zeta_n^{(2)}$, and we find, setting $\mu=\mu_t$,
\begin{subequations}    \label{eq:zeta122}
\begin{align}
    \zeta_1^{(2)}(t) &= \frac{3\na\nh}{8\pi^2 t} \ctr\, \apit
    \left[ 1 + \apit\left(
      7.43789\,\cca + 2\,\ccf - \frac{10}{9} \ctr\nf\right)\right]
       +\order{\api^3}
   \,,\label{eq:zeta122a}\\
    \zeta_2^{(2)}(t) &= \frac{\nc\nh}{4\pi^2 t}\bigg[1 + \apit
      \ccf\left(\frac{7}{4} - \frac{3}{4}\ln3\right)
      \nonumber\\&\qquad+ \apit^2\ccf\left(
        2.10889\,\cca + 2.05158\,\ccf
        - 0.268909\,\ctr\nf \right)\bigg] +\order{\api^3}\,,
        \label{eq:zeta122b}
\end{align}
\end{subequations}
where again $\apit=\api(\mu_t)$.  Since quark loops appear in
$\langle\tcalo_1(t)\rangle$ only at the two-loop level,
$\zeta_1^{(2)}(t)$ starts at $\order{\api}$.  In the case of
$\zeta_2^{(2)}(t)$, the result for general $\mu$ can be obtained through
multiplication by
\begin{equation}\label{eq:k3df1}
  \begin{split}
    \frac{m^2(\mu_t)}{m^2(\mu)} =
    1+2\,\api\,\gamma_{m,0}\,\lmut
    + 2\,\api^2\,\lmut\,\left[\gamma_{m,1}
    + \left(\frac{\beta_0\,\gamma_{m,0}}{2}
    + \gamma_{m,0}^2\right)\,\lmut\right] + \order{\api^3}\,,
  \end{split}
\end{equation}
expressing $\apit$ by $\api$ through \eqn{eq:cx3w6}, and re-expanding in
$\api$. For $\zeta_1^{(2)}(t)$ one needs to multiply by \eqn{eq:k3df1},
and in addition by $\apit/\api$.

$\zeta_{13}$ and $\zeta_{23}$ require the more sophisticated
renormalization given in \eqn{eq:zetacomp}. It is important here to work
consistently in $d$ space-time dimensions. Since $\vec{Z}_3$ contains a
$1/\ep$ pole already at $\order{\api^0}$, we need to keep the
$\order{\ep^2}$ terms of $\zeta_{2\times 2}$ at \nlo, and the
\order{\ep} terms at \nnlo. They were not required in the calculation of
\citere{Harlander:2018zpi}, so we recalculated $\zeta_{2\times 2}$,
keeping these higher terms in $\ep$. Using the identity
$\na\ctr=\nc\ccf$\cite{vanRitbergen:1998pn}, our final result for
$\vec{\zeta}_3$ reads:
\begin{subequations}\label{eq:sq2te}
\begin{align}
    \zeta_{13}(t) &= \frac{\nh \nc\ccf}{16\pi^2}\api
    \bigg\{
  5
  - 6\,\zeta(2)
  - 6\,\lmut 
  - 6\,\lmut^2 
\nonumber\\&\hspace*{1em}
  + \api\,\bigg[
      -3.31445\,\cca 
      - 27.5707\,\ccf 
      - 15.0886\,\ctr\nh 
      + 19.5780\,\ctr\nl 
\nonumber\\&\hspace*{2em}
      + \lmut\,\bigg(
          -5.68293\,\cca 
          - 32.7594\,\ccf 
          - 4.17386\,\ctr\nh 
          + 19.8261\,\ctr\nl
          \bigg)
\nonumber\\&\hspace*{2em}
      + \lmut^2\,\bigg(
          -\frac{433}{12}\,\cca 
          - \frac{33}{2}\,\ccf 
          + \frac{26}{3}\,\ctr\nf 
          \bigg) 
\nonumber\\&\hspace*{2em}
      + \lmut^3\,\bigg(
          -\frac{22}{3}\,\cca 
          - 6\,\ccf 
          + \frac{8}{3}\,\ctr\nf 
          \bigg) 
      \bigg]
  \bigg\}
        + \order{\api^3}
\,,
\label{eq:sq2tea}
\\
\zeta_{23}(t) &= \frac{\nc\nh}{2\pi^2}\bigg\{1+\lmut
    + \api\ccf\bigg[
      \frac{67}{16} 
  + \frac{13}{2}\,\ln2 
  - \frac{15}{2}\,\ln3 
  - \frac{9}{4}\,\dilogv
  \nonumber\\&\hspace{2em}
  + \bigg(
      \frac{33}{8} 
      - \ln2 
      - \frac{3}{4}\,\ln3
      \bigg)\,\lmut 
  + \frac{3}{2}\,\lmut^2 
  \bigg]
    \nonumber\\&\hspace{1em}
    + \api^2\ccf\bigg[
  -0.710509\,\cca 
  + 6.97943\,\ccf 
  - 6.43804\,\ctr\nh 
  - 2.87689\,\ctr\nl
\nonumber\\&\hspace*{2em}
  + \lmut\,\bigg(
      1.53754\,\cca 
      + 4.22899\,\ccf 
      - 4.47865\,\ctr\nh 
      - 1.47865\,\ctr\nl
      \bigg) 
\nonumber\\&\hspace*{2em}
  + \lmut^2\,\bigg(
      2.57807\,\cca 
      + 4.09934\,\ccf 
      - 0.931798\,\ctr\nf 
      \bigg) 
\nonumber\\&\hspace*{2em}
  + \lmut^3\,\bigg(
      \frac{11}{24}\,\cca 
      + \frac{3}{2}\,\ccf 
      - \frac{1}{6}\,\ctr\nf 
      \bigg) 
      \bigg]
      \bigg\}
    + \order{\api^3}\,.
\label{eq:sq2teb}
\end{align}
\end{subequations}
The logarithmic terms at \order{\api^n} are determined by the
\rg\ equation derived in \sct{sec:evol}. Nevertheless, for the
convenience of the reader, we provide the result for general $\mu$ in
this case.

This, together with the results for $\zeta_{2\times 2}$ obtained in
\citeres{Harlander:2018zpi} and explicitly given in \eqn{eq:zetaren},
completes the result for the small-flow-time coefficients of the
\ope\ up to dimension four of \eqn{eq:flowedope}.

\section{Flow-time evolution}\label{sec:evol}

In the final section of this paper we derive a general flow-time
evolution equation for flowed operators. It resembles the \rg\ equation
for regular operators but with a ``flowed anomalous dimension
matrix''. While studies of the relation between the \rg\ and the
flow-time evolution have also been performed elsewhere in the literature
(see, e.g., \citeres{Makino:2018rys,Carosso:2018rep,
  Abe:2018zdc,Carosso:2019qpb,Sonoda:2019ibh}), to our knowledge the
treatment described here has not been discussed before.

Let us return to the small-flow-time expansion of the operators $\bcalo$
defined in \eqs{eq:smalltexp}, \noeqn{eq:opoinv}, employing a matrix
rather than component-wise notation for the sake of clarity:
\begin{equation}
  \begin{split}
    \bcalo(t) = \zeta^\bare(t)\calo
    = \zeta(t)\calo^\ren\,.
    \label{eq:smallt}
  \end{split}
\end{equation}
Since we work in the small-flow-time limit, the dependence of $\zeta(t)$
on $t$ can be only through $\lmut$, defined in \eqn{eq:lmut}.  Taking
the logarithmic derivative \wrt\ $t$ of \eqn{eq:smallt}, one thus
obtains
\begin{equation}
  \begin{split}
    t\partial_t\bcalo(t) = (t\partial_t\zeta(t))\calo^\ren\,.
    \label{eq:rgttmp}
  \end{split}
\end{equation}
Using \eqn{eq:smallt} to eliminate the regular operators $\calo^\ren$,
we find the flow-equation for flowed composite operators:
\begin{equation}
  \begin{split}
    t\partial_t\bcalo(t) = \gamma^\text{f}(t)\,\bcalo(t)\,,
\quad\mbox{where}\quad
    \gamma^\text{f}(t) \equiv
    (t\partial_t\zeta(t))\zeta^{-1}(t)\,.
    \label{eq:gammaf}
  \end{split}
\end{equation}
So far the discussion is general and holds for any flowed
\ope. Specializing to our case of the \qcd\ dimension-four operators, we
can write the ``flowed anomalous dimension'' matrix as
\begin{equation}\label{eq:qldnp}
  \begin{split}
    \gamma^\text{f} =
    \left(
    \begin{matrix}
      \gamma^\text{f}_{2\times 2} & \vec{\gamma}^\text{f}_3
      \\
      0 & 0
    \end{matrix}
    \right)\,,\qquad
    \gamma^\text{f}_{2\times 2}(t) &=
      \left(t\partial_t\zeta_{2\times
        2}(t)\right)\zeta^{-1}_{2\times 2}(t)\,,\\
    \vec{\gamma}^\text{f}_3(t) &= -\gamma^\text{f}_{2\times 2}(t) \vec{\zeta}_3(t) +
      t\partial_t\vec{\zeta}_3(t)\,.
  \end{split}
\end{equation}
Through \order{\api^2}, the result can be directly evaluated from
\eqs{eq:zetaren} and \noeqn{eq:sq2te}. A consistency check is obtained
by noting that $\zeta(t)$ depends on $t$ only through $\lmut$:
\begin{equation}\label{eq:sr6yc}
  \begin{split}
    t\partial_t\zeta(t) =
    \mu^2\deriv{}{}{\mu^2}\zeta(t) =
    \mu^2\dderiv{}{}{\mu^2}\zeta(t) -
    \api\beta\deriv{}{}{\api}\zeta(t)\,.
  \end{split}
\end{equation}
On the other hand, we know that $\api\tcalo_1(t)$ and $\tcalo_2(t)$ are
\rg\ invariant\cite{Luscher:2010iy,Luscher:2011bx,Makino:2014taa} and
therefore, with \eqn{eq:smalltexp},
\begin{equation}\label{eq:9q9qx}
  \begin{split}
    \left(
    \begin{matrix}
      0\\0\\4m^4\gamma_m
    \end{matrix}
    \right) &= \mu^2\dderiv{}{}{\mu^2}H^{-1}(\api)\tcalo(t)=\\&
    = \mu^2\dderiv{}{}{\mu^2}H^{-1}(\api)
    \left(\zeta^{(0)}(t)\mathds{1} + \zeta^{(2)}(t)m^2 \mathds{1}
    + \zeta(t)\calo^\ren\right)\,,
  \end{split}
\end{equation}
where
\begin{equation}\label{eq:rdi24}
  \begin{split}
    H(x) = \left(
    \begin{matrix}
      H_{2\times 2}(x) &\vec{0}\\
      \vec{0}^{\,T} & 1
    \end{matrix}
    \right)\,,\quad\text{with}\quad
        H_{2\times 2}(x) = \left(
    \begin{matrix}
      (4\pi^2 x)^{-1} & 0 \\
      0 & 1
    \end{matrix}
    \right)\,.
  \end{split}
\end{equation}
Since operators of different mass dimensions do not mix under
\rg\ evolution and $\zeta_3^{(0,2)}(t)=0$, we can drop the first two terms
in the brackets on the r.h.s.\ of \eqn{eq:9q9qx}.\footnote{This can also
  be seen by noting that these two terms, multiplied by $H^{-1}(\api)$,
  are $\api\langle\tcalo_1(t)\rangle$ and $\langle\tcalo_2(t)\rangle$,
  expanded through order $m^2$.}  We thus arrive at
\begin{equation}\label{eq:mudmuzeta}
  \begin{split}
    \mu^2\dderiv{}{}{\mu^2}\zeta(t)=
    \left(
    \begin{matrix}
      0 & 0 & 0\\
      0 & 0 & 0\\
      0 & 0 & 4\gamma_m
    \end{matrix}
    \right)
    -
    \left(
    \begin{matrix}
      \beta & 0 &0 \\ 0 & 0 &0\\0&0&0
    \end{matrix}
    \right)\zeta(t) - \zeta(t)\gamma^\calo\,,
  \end{split}
\end{equation}
where $\gamma^\calo$ is the anomalous dimension of the operators
$\calo^\ren$, defined through
\begin{equation}\label{eq:hdahz}
  \begin{split}
    \mu^2\dderiv{}{}{\mu^2}\calo^\ren = \gamma^\calo(\api)\calo^\ren\,.
  \end{split}
\end{equation}
It can be written as
\begin{equation}\label{eq:uo5rb}
  \begin{split}
    \gamma^\calo = \left(\mu^2\dderiv{}{}{\mu^2}Z\right)Z^{-1} = \left(
    \begin{matrix}
      \gamma^\calo_{2\times 2} & \vec{\gamma}^{\,\calo}_3\\[.5em]
      \vec{0}^{\,T} & 4\gamma_m
    \end{matrix}
    \right)\,,
  \end{split}
\end{equation}
with $Z$ from \eqn{eq:z}.  Using the expressions of \sct{sec:ope}, one
derives\cite{Spiridonov:1984br,Spiridonov:1988md}
\begin{equation}\label{eq:mzw19}
  \begin{split}
    \gamma^\calo_{2\times 2} &= \left(\mu^2\dderiv{}{}{\mu^2} Z_{2\times
      2}\right)Z^{-1}_{2\times 2} = \left(
    \begin{matrix}
      -\displaystyle \api\deriv{}{}{\api}\beta
      & \displaystyle -2\,\api\deriv{}{}{\api}\gamma_m\\[1em]
      0 & 0
    \end{matrix}
    \right)\,,\\
    \vec{\gamma}^{\,\calo}_3 &= Z_m^4\left(\mu^2\dderiv{}{}{\mu^2}\vec{Z}_3 -
    \gamma^\calo_{2\times 2}\vec{Z}_3\right) =
    \left(
    \begin{matrix}
      \displaystyle 4\api\deriv{}{}{\api}\gamma_0\\[1em]
      8\gamma_0
    \end{matrix}
    \right)\,.
  \end{split}
\end{equation}
The \qcd\ renormalization group functions $\beta$ and $\gamma_m$ are
defined in \eqs{eq:betaep} and \noeqn{eq:gammam}, respectively.  Since
they are of \order{\api}, the explicit $\mu$-dependence of
$\zeta_{2\times 2}(t)$ can be derived through \order{\api^3} from the
results of \citere{Harlander:2018zpi}. Thus, for
$\gamma^\text{f}_{2\times 2}$, \eqn{eq:sr6yc} is not just a consistency
check, but a means to derive higher order terms. In our case, we can
obtain the result through \order{\api^3}:
\begin{subequations}    \label{eq:gamma}
\begin{align}
    \gamma^\text{f}_{11} &=
 \api^2\,\bigg[
      \frac{3}{32}\,\cca^2 
      + \frac{1}{8}\,\cca\ctr\nf 
      + \frac{7}{8}\,\ccf\ctr\nf
      \bigg] 
  + \api^3\,\bigg[
      -\frac{7}{48}\,\cca\ctr^2\nf^2 
\nonumber\\&\hspace*{1em}
      - \frac{35}{36}\,\ccf\ctr^2\nf^2 
      + \cca\ccf\ctr\nf\,\bigg(
          \frac{11891}{2880} 
          + \frac{27}{40}\,\ln2 
          - \frac{81}{160}\,\ln3
          \bigg) 
\nonumber\\&\hspace*{1em}
      + \cca^2\ctr\nf\,\bigg(
          \frac{1687}{2880} 
          + \frac{29}{40}\,\ln2 
          - \frac{9}{20}\,\ln3
          \bigg) 
      + \cca^3\,\bigg(
          \frac{6643}{11520} 
          - \frac{319}{160}\,\ln2 
          + \frac{99}{80}\,\ln3
          \bigg) 
\nonumber\\&\hspace*{1em}
      + \ccf^2\ctr\nf\,\bigg(
          \frac{25}{64} 
          + \frac{45}{8}\,\ln2 
          - \frac{111}{32}\,\ln3 
          + \frac{3}{4}\,\dilogv 
          - \frac{3}{8}\,\zeta(2)
          \bigg)
\nonumber\\&\hspace*{1em}
      + \bigg(
          \frac{1}{6}\,\cca^2\ctr\nf 
          + \frac{11}{64}\,\cca^3 
          + \frac{77}{48}\,\cca\ccf\ctr\nf 
          - \frac{1}{12}\,\cca\ctr^2\nf^2 
          - \frac{7}{12}\,\ccf\ctr^2\nf^2
          \bigg)\,\lmut 
          \bigg]
\nonumber\\&\hspace*{1em}
  + \order{\api^4}\,,
  \\
  \gamma^\text{f}_{12} &=
  -\frac{3}{2}\,\api\,\ccf 
  + \api^2\,\Bigg\{
      -\frac{367}{48}\,\cca\ccf 
      + \frac{5}{3}\,\ccf\ctr\nf 
      + \ccf^2\,\bigg[
          -\frac{3}{16} 
          - \frac{3}{2}\,\ln2
          - \frac{9}{8}\,\ln3
          \bigg] 
\nonumber\\&\hspace*{1em}
      + \bigg[
          -\frac{11}{4}\,\cca\ccf 
          + \ccf\ctr\nf
          \bigg]\,\lmut
      \Bigg\} 
\nonumber\\&\hspace*{0em}
  + \api^3\,\Bigg\{
      \ccf^2\cca\,\bigg[
          -\frac{391}{768} 
          - \frac{431}{24}\,\ln2 
          + \frac{3}{8}\,\ln^22 
          + \frac{11}{64}\,\ln3 
          + \frac{9}{8}\,\dilogv 
          - \frac{33}{64}\,\zeta(2)
          \bigg] 
\nonumber\\&\hspace*{1em}
      + \ccf\ctr^2\nf^2\,\bigg[
          -\frac{25}{18} 
          - \frac{1}{2}\,\zeta(2)
          \bigg] 
      + \ccf^3\,\bigg[
          -\frac{1401}{256} 
          + \frac{339}{16}\,\ln2 
          - \frac{9}{8}\,\ln^22 
\nonumber\\&\hspace*{2em}
          - \frac{657}{64}\,\ln3 
          - \frac{9}{4}\,\ln2\,\ln3 
          - \frac{27}{32}\,\ln^23 
          + \frac{153}{32}\,\dilogv 
          + \frac{39}{64}\,\zeta(2)
          \bigg] 
\nonumber\\&\hspace*{1em}
      + \cca^2\ccf\,\bigg[
          -\frac{5291}{144} 
          - \frac{2311}{24}\,\ln2 
          + \frac{4641}{64}\,\ln3 
          + \frac{11}{32}\,\dilogv 
          + \frac{11}{8}\,\zeta(2)
          \bigg] 
\nonumber\\&\hspace*{1em}
      + \ccf^2\ctr\nf\,\bigg[
          \frac{8827}{960} 
          - \frac{2089}{120}\,\ln2 
          + \frac{847}{80}\,\ln3 
          + \frac{3}{8}\,\dilogv 
          - \frac{3}{16}\,\zeta(2) 
          - \frac{9}{2}\,\zeta(3)
          \bigg] 
\nonumber\\&\hspace*{1em}
      + \cca\ccf\ctr\nf\,\bigg[
          \frac{5861}{360} 
          + \frac{4273}{120}\,\ln2 
          - \frac{2139}{80}\,\ln3 
          - \frac{1}{8}\,\dilogv 
          + \frac{7}{8}\,\zeta(2) 
          + \frac{9}{2}\,\zeta(3)
          \bigg] 
\nonumber\\&\hspace*{2em}
      + \frac{3}{32}\,\ccf\,c_\chi^{(2)}
      + \bigg[
          -\frac{4445}{192}\,\cca^2\ccf 
          + \frac{647}{48}\,\cca\ccf\ctr\nf 
          - \frac{5}{3}\,\ccf\ctr^2\nf^2 
\nonumber\\&\hspace*{2em}
          + \ccf^2\cca\,\bigg(
              -\frac{33}{64} 
              - \frac{33}{8}\,\ln2 
              - \frac{99}{32}\,\ln3
              \bigg) 
          + \ccf^2\ctr\nf\,\bigg(
              \frac{15}{16} 
              + \frac{3}{2}\,\ln2 
              + \frac{9}{8}\,\ln3
              \bigg)
          \bigg]\,\lmut 
\nonumber\\&\hspace*{1em}
      + \bigg[
          -\frac{121}{32}\,\cca^2\ccf 
          + \frac{11}{4}\,\cca\ccf\ctr\nf 
          - \frac{1}{2}\,\ccf\ctr^2\nf^2
          \bigg]\,\lmut^2 
      \Bigg\}
        + \order{\api^4}\,,\\
        \gamma^\text{f}_{21} &=
        \api^3\,\bigg[
      \cca^2\ctr\nf\,\bigg(
          \frac{599}{5760} 
          + \frac{33}{80}\,\ln2 
          - \frac{99}{320}\,\ln3
          \bigg) 
      + \cca\ctr^2\nf^2\,\bigg(
          \frac{41}{1440} 
\nonumber\\&\hspace*{2em}
          - \frac{3}{20}\,\ln2 
          + \frac{9}{80}\,\ln3
          \bigg) 
      + \cca\ccf\ctr\nf\,\bigg(
          \frac{209}{1152} 
          + \frac{55}{16}\,\ln2 
          - \frac{407}{192}\,\ln3 
\nonumber\\&\hspace*{2em}
          + \frac{11}{24}\,\dilogv 
          - \frac{11}{48}\,\zeta(2)
          \bigg) 
      + \ccf\ctr^2\nf^2\,\bigg(
          \frac{43}{144} 
          - \frac{5}{4}\,\ln2 
          + \frac{37}{48}\,\ln3 
\nonumber\\&\hspace*{2em}
          - \frac{1}{6}\,\dilogv 
          + \frac{1}{12}\,\zeta(2)
          \bigg)
          \bigg]
  + \order{\api^4}\,,\\
  \gamma^\text{f}_{22} &=
\api^2\,\Bigg[
      \cca\ccf\,\bigg(
          \frac{11}{96} 
          - \frac{11}{12}\,\ln2 
          - \frac{11}{16}\,\ln3
          \bigg)
      + \ccf\ctr\nf\,\bigg(
          -\frac{2}{3} 
          + \frac{1}{3}\,\ln2 
\nonumber\\&\hspace*{2em}
          + \frac{1}{4}\,\ln3
          \bigg)
      \Bigg]
  + \api^3\,\Bigg\{
      \cca\ccf\ctr\nf\,\bigg[
          \frac{2521}{960} 
          - \frac{1709}{90}\,\ln2 
          - \frac{1}{6}\,\ln^22 
\nonumber\\&\hspace*{2em}
          + \frac{1537}{160}\,\ln3 
          - \frac{1}{24}\,\dilogv
          \bigg] 
      + \cca^2\ccf\,\bigg[
          -\frac{7397}{2304} 
          - \frac{213}{16}\,\ln2 
          + \frac{11}{24}\,\ln^22 
\nonumber\\&\hspace*{2em}
          + \frac{107}{16}\,\ln3 
          + \frac{11}{8}\,\dilogv 
          - \frac{121}{192}\,\zeta(2)
          \bigg] 
      + \ccf\ctr^2\nf^2\,\bigg[
          -\frac{749}{720} 
          + \frac{763}{90}\,\ln2 
\nonumber\\&\hspace*{2em}
          - \frac{83}{20}\,\ln3 
          - \frac{1}{6}\,\dilogv 
          + \frac{1}{12}\,\zeta(2)
          \bigg] 
      + \ccf^2\ctr\nf\,\bigg[
          -\frac{139}{192} 
          - \frac{119}{8}\,\ln2 
\nonumber\\&\hspace*{2em}
          + \frac{1}{6}\,\ln^22 
          + \frac{261}{32}\,\ln3 
          + \frac{1}{2}\,\ln2\,\ln3 
          + \frac{3}{16}\,\ln^23 
          - \frac{23}{8}\,\dilogv 
          + \frac{5}{48}\,\zeta(2)
          \bigg] 
\nonumber\\&\hspace*{1em}
      + \ccf^2\cca\,\bigg[
          \frac{253}{384} 
          + \frac{209}{8}\,\ln2 
          - \frac{11}{24}\,\ln^22 
          - \frac{99}{8}\,\ln3 
          - \frac{11}{8}\,\ln2\,\ln3 
          - \frac{33}{64}\,\ln^23 
\nonumber\\&\hspace*{2em}
          + \frac{187}{32}\,\dilogv 
          + \frac{143}{192}\,\zeta(2)
          \bigg] 
      + \bigg(\frac{11}{96}\,\cca
      - \frac{1}{24}\,\ctr\nf\bigg)\,c_\chi^{(2)}
\nonumber\\&\hspace*{2em}
      + \bigg[
          \cca^2\ccf\,\bigg(
              \frac{121}{576} 
              - \frac{121}{72}\,\ln2 
              - \frac{121}{96}\,\ln3
              \bigg) 
          + \ccf\ctr^2\nf^2\,\bigg(
              \frac{4}{9} 
              - \frac{2}{9}\,\ln2 
              - \frac{1}{6}\,\ln3
              \bigg) 
\nonumber\\&\hspace*{2em}
          + \cca\ccf\ctr\nf\,\bigg(
              -\frac{187}{144} 
              + \frac{11}{9}\,\ln2 
              + \frac{11}{12}\,\ln3
              \bigg)
          \bigg]\,\lmut 
      \Bigg\}
        + \order{\api^4}\,.
\end{align}
\end{subequations}
We verified that this agrees through $\order{\api^2}$ with the result
which is obtained by directly inserting \eqn{eq:zetaren} into
\eqn{eq:qldnp}. Due to the factor of $1/g^2$ in $\tcalo_1$ (see
\eqn{eq:flopo}), $\gamma^\text{f}_{2\times 2}$ is not \rg\ invariant,
while $H^{-1}_{2\times 2}\gamma^\text{f}_{2\times 2} H_{2\times
  2}$ is. It may be useful to note that, by subtracting the \vev{}s
off of $\tcalo_1$ and $\tcalo_2$,
\begin{equation}\label{eq:oa0sw}
  \begin{split}
    \tcalo_{1,\text{sub}}(t,x) &= \tcalo_1(t,x) - \langle\tcalo_1(t,x)\rangle\,,\\
    \tcalo_{2,\text{sub}}(t,x) &= \tcalo_2(t,x) - \langle\tcalo_2(t,x)\rangle\,,
  \end{split}
\end{equation}
the resulting operators do not mix with $\tcalo_3$ under
$t$-evolution. Rather, their logarithmic $t$-evolution is fully governed
by $\gamma^\text{f}_{2\times 2}$ and thus known through $\order{\api^3}$.

\eqn{eq:mudmuzeta} does not analogously allow one to derive the
\order{\api^3} terms of $\vec{\gamma}^\text{f}_3$, because it involves
$\gamma_0$ which, in contrast to $\beta$ and $\gamma_m$, starts at
$\order{\api^0}$ rather than $\order{\api}$, see
\eqn{eq:gamma0}. Therefore, we can only give the result through
\order{\api^2} for $\vec{\gamma}^\text{f}_3$:
\begin{subequations}\label{eq:gammaf3}
  \begin{align}
  \gamma_{13}^\text{f} &=
  \frac{3\nc\nh}{8\pi^2}\api\ccf
  \bigg\{ 1 +
  \api\bigg[
  9.24729\,\cca 
  - 2.47340\,\ccf 
\nonumber\\&\hspace*{1em}
  - 2.91787\,\ctr\nh 
  + 1.08213\,\ctr\nl
  + \lmut\,\bigg(
      \frac{11}{6}\,\cca 
      + 3\,\ccf 
      - \frac{2}{3}\,\ctr\nf 
      \bigg) 
      \bigg]\bigg\}\,,\\
  \gamma_{23}^\text{f} &=
  \frac{\nc\nh}{2\pi^2}
  \bigg\{
  1 + \api\ccf\left[
    \frac{33}{8}
    - \ln2
    - \frac{3}{4}\ln3
    + 3\,\lmut
    \right]
  +\api^2\ccf\bigg[
  2.81363\,\cca 
  + 4.22899\,\ccf 
\nonumber\\&\hspace*{1em}
  - 4.31769\,\ctr\nh 
  - 1.31769\,\ctr\nl
  + \lmut\,\bigg(
      6.43224\,\cca 
      + 8.19868\,\ccf 
      - 1.70263\,\ctr\nf 
      \bigg) 
\nonumber\\&\hspace*{1em}
  + \lmut^2\,\bigg(
      \frac{11}{8}\,\cca 
      + \frac{9}{2}\,\ccf 
      - \frac{1}{2}\,\ctr\nf 
      \bigg)
    \bigg]
  \bigg\}\,.
  \end{align}
\end{subequations}
We checked, of course,
that \eqn{eq:mudmuzeta} is consistent with the results for $\zeta_{13}$
and $\zeta_{23}$ of \eqn{eq:sq2te}.

\section{Conclusions}\label{sec:conclusions}

We presented the flowed \ope{} for general current correlators and its
matching to regular \qcd{} through \nnlo{} in the strong coupling
$\alpha_s$ and through mass dimension four by using the small-flow-time
expansion.  Our calculation is based on the renormalization procedure
for the regular \qcd{} dimension-four operators worked out in
\citere{Spiridonov:1984br,Spiridonov:1988md}, the mixing matrix between
flowed and regular operators derived in \citere{Harlander:2018zpi}, the
method of projectors \cite{Gorishnii:1983su}, and the tools and results
for perturbative calculations in the \gff\ presented in
\citere{Artz:2019bpr}.

Overall, our results allow to combine the known perturbative results for
the regular \qcd\ current correlators from the literature to
gradient-flow lattice calculations. This lays out the path for an
alternative determination of hadronic contributions to observables such
as the anomalous magnetic moment of the muon.  In addition, we derived a
general logarithmic flow-time evolution equation for flowed operators
and presented its explicit form for the dimension-four operators
considered in this paper.

Our methods are sufficiently general to be applied to similar problems
at higher orders in perturbation theory, such as \abbrev{CP} violating
operators \cite{Rizik:2020naq} relevant for the electric dipole moment
of the neutron, or four-quark operators occurring in flavor
physics\cite{Suzuki:2020zue}.

\paragraph{Acknowledgments.}
We would like to thank A.~Hoang for confirmative comments on the
validity of the approach pursued in this paper, and Y.~Kluth for helpful
comments and clarifications. \abbrev{RVH} would like to thank
K.~Chetyrkin for past initiation and collaboration on the material of
\sct{sec:ccc}.

This work was supported by \textit{Deutsche
  Forschungsgemeinschaft (DFG)} through project
\href{http://gepris.dfg.de/gepris/projekt/386986591}{HA 2990/9-1}, and
by the U.S.\ Department of Energy under award No.\ \abbrev{DE-SC0008347}.  This
document was prepared using the resources of the Fermi National
Accelerator Laboratory (Fermilab), a U.S. Department of Energy, Office
of Science, HEP User Facility. Fermilab is managed by Fermi Research
Alliance, LLC (FRA), acting under Contract No.\ DE-AC02-07CH11359.

\begin{appendix}

\section{Renormalization group functions}\label{app:rg}

The $d$-dimensional beta function is defined as
\begin{equation}\label{eq:v6pj6}
  \begin{split}
    \mu^2\dderiv{}{}{\mu^2}a_s(\mu) = \api(\mu)\beta_\ep(\api(\mu))\,,
  \end{split}
\end{equation}
where $\api\equiv \alpha_s/\pi \equiv g^2/(4\pi^2)$. The renormalized
coupling $g=g(\mu)$ is related to the bare one through $g_\bare =
\hat\mu^{\ep}\,Z_s^{1/2}\,g$, where $Z_s$ is the \msbar\ renormalization
constant. From this follow the relations
\begin{equation}
  \begin{split}
    \beta_\ep(\api) &=
 -\ep\left(1+\api\deriv{}{}{\api}\ln Z_s(\api)\right)^{-1} =
 -\ep +\beta(\api) \equiv -\ep- \sum_{n\geq 0}\api^{n+1}\beta_n\,,\\
     Z_s(\api) &= 1
    -\frac{\api}{\ep}\beta_0 + \api^2\left(\frac{1}{\ep^2}\beta_0^2
    -\frac{1}{2\ep}\beta_1\right) + \order{\api^3}\,.
    \label{eq:betaep}
  \end{split}
\end{equation}
Through \sct{sec:evol}, we only need the first two perturbative
coefficients, while $\beta_2$ is required in order to derive the
$\order{\api^3}$ terms of $\gamma^\text{f}$ in \eqn{eq:gamma}:
\begin{equation}
  \begin{split}
    \beta_0 &=
      \frac{1}{4}\left(\frac{11}{3}\cca -
      \frac{4}{3}\ctr\nf\right)\,,\qquad \beta_1 =
      \frac{1}{16}\left(\frac{34}{3}\cca^2
      -4\ccf\ctr\nf-\frac{20}{3}\cca\ctr\nf\right)\,,\\
 \beta_{2} & =  \frac{1}{64}\bigg(\frac{2857}{54} \cca^3
 +2 \ccf^2 \ctr\nf - \frac{205}{9} \ccf\cca\ctr\nf 
   - \frac{1415}{27} \cca^2 \ctr \nf
\\&\qquad
   + \frac{44}{9} \ccf \ctr^2 \nf^2
  + \frac{158}{27} \cca \ctr^2 \nf^2  \bigg)\,.
    \label{eq:beta}
  \end{split}
\end{equation}
The anomalous dimension of the quark mass is defined through
\begin{equation}
  \begin{split}
    \gamma_m(\api) &= -\api\beta_\ep(\api)\deriv{}{}{\api}\ln Z_m(\api) \equiv
    -\sum_{n\geq 0} \api^{n+1}\gamma_{m,n}\,,
    \label{eq:gammam}
  \end{split}
\end{equation}
with the first three perturbative coefficients given by
\begin{equation}
  \begin{split}
    \gamma_{m,0} &= \frac{3}{4}\ccf\,,\qquad
    \gamma_{m,1} = \frac{3}{32}\ccf^2 + \frac{97}{96}\cca\ccf
    - \frac{5}{24}\ccf\ctr\nf\,,\\
 \gamma_{m,2} & = \frac{1}{64} \left[
  \frac{129}{2} \ccf^3 - \frac{129}{4}\ccf^2 \cca
 + \frac{11413}{108}\ccf \cca^2 \right. \\ & \left.
 +\ccf^2 \ctr \nf (-46+48\zeta(3))
+\ccf\cca\ctr\nf \left( -\frac{556}{27}-48\zeta(3) \right)  
- \frac{140}{27} \ccf\ctr^2 \nf^2  \right]  \,.
  \end{split}
\end{equation}
It determines the \msbar\ renormalized mass $m$ through
\begin{equation}\label{eq:g3kpi}
  \begin{split}
    m_\bare = Z_m\,m\,,\qquad
    Z_m = 1 -\frac{\api}{\ep} \gamma_{m,0}
    + \api^2\left[\frac{\gamma_{m,0}}{2\ep^2}\left(
      \gamma_{m,0}+\beta_0\right) - \frac{1}{2\ep}\gamma_{m,1}\right]
    + \order{\api^3}\,.
  \end{split}
\end{equation}
Similarly to $\beta_\ep$, the third coefficient $\gamma_{m,2}$ is needed
only in \sct{sec:evol}.

The renormalization constant of the vacuum energy $Z_0$ is related to
the corresponding anomalous dimension $\gamma_0$ through
\begin{equation}\label{eq:gamma0}
  \begin{split}
    \gamma_0(\api) &= \left[4\gamma_m(\api)-\ep\right]Z_0(\api)
    +\beta_\ep(\api)\api\deriv{}{}{\api} Z_0(\api)
    \equiv -\frac{\nc\nh}{(4\pi)^2}\sum_{n\geq 0}\api^n\gamma_{0,n}\,,
  \end{split}
\end{equation}
which leads to
\begin{equation}\label{eq:z0}
  \begin{split}
    Z_0(\api) &= \frac{\nc\nh}{(4\pi)^2\ep}
    \bigg\{1
      + \api\left(
       \frac{\gamma_{0,1}}{2}
      -\frac{2\gamma_{m,0}}{\ep}
      \right)
      + \api^2\Big[
      \frac{2}{3\ep^2}\,\left(
      \beta_0\,\gamma_{m,0}
      + 4\,\gamma_{m,0}^2\right)\\&\qquad
      - \frac{1}{6\ep}\left(
       \beta_0\,\gamma_{0,1}
      + 4\,\gamma_{0,1}\gamma_{m,0}
      + 8\,\gamma_{m,1}
      \right)
      + \frac{1}{3}\gamma_{0,2}\Big]\bigg\}
       + \order{\api^3}\,.
  \end{split}
\end{equation}
The first three perturbative coefficients are given by
\cite{Spiridonov:1988md,Chetyrkin:1994ex}\footnote{Higher orders have
  been computed in \citere{Chetyrkin:2018avf}.}
\begin{equation}
  \begin{split}
    \gamma_{0,0} &= 1\,,\qquad \gamma_{0,1} = \ccf\,,\\
    \gamma_{0,2} &= -\ccf^2\,\left(\frac{131}{32}-3\,\zeta(3)\right)
    - \ccf\cca\left(-\frac{109}{32}+\frac{3}{2}\,\zeta(3)\right)
    - \ccf\ctr\left(\frac{5}{8}\nf+3\nh\right)\,.
    \label{eq:gamma0coef}
  \end{split}
\end{equation}

\section{Perturbative coefficient functions}\label{app:coefs}

This appendix cites the results for the coefficient functions $C_n$ of
the regular dimension-four operators appearing in the \ope\ of the current
correlators defined in \eqn{eq:ccope}. We consider scalar,
pseudo-scalar, vector- and axial-vector currents, both diagonal and
non-diagonal, i.e.\ the currents assume the form
\begin{equation}\label{eq:currents}
  \begin{split}
    j(x) &= \bar\psi_k(x)\Gamma\psi_l(x)\,,\qquad
    \Gamma\in\{1,i\gamma_5,\gamma^\mu,\gamma^\mu\gamma_5\}\,,\\ k,l &\in
    N\cup M\,,\quad M=\{1,\ldots,\nh\}\,,\quad N=\{\nh+1,\ldots,\nf\}\,.
  \end{split}
\end{equation} 
This means that $\psi_k$ and $\psi_l$ can be either both massive with
mass $m$ ($k,l\in M$), or both massless ($k,l\in N$), or one of
them is massless, the other massive (e.g.\ $k\in M$, $l\in N$).  While
$C_1$ is independent of $k$ and $l$, the coefficient $C_2$ of the quark
operator takes the form
\begin{equation}\label{eq:e12a9}
  \begin{split}
    C_2 = C_{2,N} + \frac{1}{\nh}\left(\delta_{kM}+\delta_{lM}\right)
    \left( C_{2,M} + C_{2,\text{nd}} \right)\,,
  \end{split}
\end{equation}
where $\delta_{kM}=1$ for $k\in M$, and 0 otherwise. Also the results
for $C_3$ depend on whether the quarks $k$ and $l$ are massive or
not. This dependence will be indicated explicitly below, using the
$\delta_{kM}$ symbol defined above.

For convenience, we introduce the short-hand notation
\begin{equation}
  \begin{split}
    \logmuqms \equiv \ln\frac{Q^2}{\mu^2}\,,
    \label{eq:lqdef}
  \end{split}
\end{equation}
and the dimensionless quantities
\begin{equation}
  \hat C_1 = Q^4\,C_1\,,\qquad
  \hat C_2 = -2\,Q^4\,C_2\,,\qquad
  \hat C_3 = Q^4\,C_3\,.
\end{equation}
The extra factor $(-2)$ between $C_2$ and $\hat C_2$ arises from
using $\calo_2'$ in \citere{Harlander:diss} rather than $\calo_2$ from
\eqn{eq:calo}. For the sake of brevity, we insert the SU(3) color factors.

\subsection{Vector and axial-vector currents}

In this case, the current correlator can be decomposed into a
transversal and a longitudinal part, each associated with a set of
coefficient functions:
\begin{equation}
\int\dd^4 xe^{iqx}j^{v/a}_\mu(x)j^{v/a}_\nu(0) \stackrel{Q^2\to
  \infty}{\longrightarrow} \sum_n \left((g_{\mu\nu}Q^2 - Q_\mu Q_\nu)
  C_n^{v/a,{\rm T}} - Q_\mu Q_\nu C_n^{v/a,{\rm L}}\right) \calo^\ren_n\,.
\end{equation}
The upper sign refers to the vector, the lower sign to the axial-vector
case. The results have been taken from \citere{Harlander:diss}:
\begin{subequations}
\begin{align}
  &       \hat C^{v/a,{\rm T}}_1 = \frac{1}{12}\,\api
  + \frac{7}{72}\,\api^2\,,
  \label{c1vat}
  \\[.5em]
  & \hat C^{v/a,{\rm T}}_{2,N} = \api^2
  \left(
  -1 
  + \frac{1}{3}\,\logmuqms
  + \frac{4}{3}\,\zeta(3)
  \right)\,,
  \label{c2ssval}
  \\[.5em]
  & \hat C^{v/a,{\rm T}}_{2,M} = 
  - \api
  + \api^2\,\bigg[
    -\frac{29}{6} 
    + \frac{1}{6}\,\nf
    + \logmuqms\,\bigg(
    -\frac{11}{4} 
    + \frac{1}{6}\,\nf
    \bigg) 
    \bigg]\,,\\[.5em]
  & \hat C^{v/a,{\rm T}}_{2,\text{nd}} = 
  \pm\left(1 
  + {4\over 3}\,\api
  + \api^2\,\bigg[
    {191\over 18} 
    - {7\over 27}\,\nf
    + \logmuqms\,\bigg(
    {11\over 3} 
    - {2\over 9}\,\nf
    \bigg) 
    \bigg]\right)\,,\\[.5em]
  & \hat C_1^{v/a,{\rm L}} = 0\,,
  \qquad \hat C_{2,N}^{v/a,{\rm L}} = 0\,,
  \qquad \hat C_{2,M}^{v/a,{\rm L}} = 1\,,
  \qquad \hat C_{2,\text{nd}}^{v/a,{\rm L}} = \mp 1\,,
  \\[.5em]
  & \hat C^{v/a,{\rm T}}_3 =  
  \frac{3}{16\pi^2}\,\bigg\{ \delta_{kM}\delta_{lM}\,\Bigg[
    \api\,\bigg(
    \frac{152}{9} 
    - \frac{32}{3}\,\zeta(3)
    \bigg)
    \nonumber\\&\mbox{\hspace{2em}}
    + \api^2\,\bigg[
      \frac{1295}{9} 
      - \frac{524}{3}\,\zeta(3) 
      + 120\,\zeta(5)
      + \nf\,\bigg(
      -\frac{362}{81} 
      + \frac{8}{27}\,\zeta(3)
      \bigg) 
      \nonumber\\&\mbox{\hspace{3em}} 
      + \logmuqms\,\bigg(
      114 
      - 72\,\zeta(3) 
      + \nf\,\left(
      -\frac{76}{27} 
      + \frac{16}{9}\,\zeta(3)
      \right)
      \bigg) 
      \bigg]
    \Bigg]
  \nonumber\\&\mbox{\hspace{1em}} 
  \pm 2\delta_{kM}\delta_{lM}\,\Bigg[
    -4\,\logmuqms 
    + \api\,\bigg(
    -\frac{56}{3} 
    + 16\,\zeta(3)
    - \frac{32}{3}\,\logmuqms 
    - 8\,\logmuqms^2 
               \bigg)
\nonumber\\&\mbox{\hspace{2em}} 
                + \api^2\,\bigg[
               -\frac{18967}{108} 
               + \frac{4588}{27}\,\zeta(3) 
               + \frac{4}{3}\,\zeta(4) 
               - \frac{280}{27}\,\zeta(5)
               + \nf\,\bigg(
                   \frac{337}{54} 
                   - \frac{40}{9}\,\zeta(3)
                   \bigg) 
\nonumber\\&\mbox{\hspace{3em}} 
                + \logmuqms\,\bigg(
                   -\frac{3617}{18} 
                   + \frac{332}{3}\,\zeta(3)
                   + \nf\,\left(
                       \frac{157}{27} 
                       - \frac{8}{3}\,\zeta(3)
                       \right)
                       \bigg)
\nonumber\\&\mbox{\hspace{3em}}
                + \logmuqms^2\,\bigg(
                   -77 
                   + \frac{22}{9}\,\nf
                   \bigg) 
                + \logmuqms^3\,\bigg(
                   -18 
                   + \frac{4}{9}\,\nf
                   \bigg) 
               \bigg]
           \Bigg]
\nonumber\\&\mbox{\hspace{1em}} 
       + ( \delta_{kM} + \delta_{lM} ) \, \Bigg[
           -2 
                + \api\,\bigg(
               -4 
               - 4\,\logmuqms
               \bigg)
\nonumber\\&\mbox{\hspace{2em}} 
                + \api^2\,\bigg[
               -\frac{776}{27} 
               + \frac{2996}{27}\,\zeta(3) 
               - \frac{3440}{27}\,\zeta(5)
               + \nf\,\bigg(
                   \frac{23}{27} 
                   - \frac{4}{9}\,\zeta(3)
                   \bigg) 
\nonumber\\&\mbox{\hspace{3em}} 
                + \logmuqms\,\bigg(
                   -36 
                   + \frac{10}{9}\,\nf
                   \bigg) 
               - 8\,\logmuqms^2 
               \bigg]
           \Bigg]
\nonumber\\&\mbox{\hspace{1em}} 
       + \nh \,\api^2\,\Bigg[
           -\frac{80}{9} 
           + \frac{16}{3}\,\zeta(3)
                + \logmuqms\,\bigg(
               \frac{40}{9} 
               - \frac{16}{3}\,\zeta(3)
               \bigg)
           - \frac{2}{3}\,\logmuqms^2 
           \Bigg]
\nonumber\\&\mbox{\hspace{1em}} 
 + \nh \,\api^2\,\Bigg[
          ( \delta_{kM} + \delta_{lM} )\,\bigg(
               \frac{100}{9} 
               - \frac{16}{3}\,\zeta(3)
               \bigg) 
\nonumber\\&\mbox{\hspace{1em}} 
           \pm \delta_{kM}\delta_{lM}\,\bigg(
               -\frac{64}{9} 
               + 16\,\logmuqms 
               + \frac{32}{3}\,\zeta(3)
               \bigg)
           \Bigg]
\bigg\}\,,
\\[.5em]
& \hat C_3^{v/a,{\rm L}} = \frac{3}{16\pi^2}\,\bigg\{\delta_{kM}\delta_{lM}\,\Bigg[
           8  + \api\,\bigg(
               \frac{152}{3} 
               + 32\,\logmuqms
               \bigg)
\nonumber\\&\mbox{\hspace{2em}} 
                + \api^2\,\bigg[
               \frac{7306}{9} 
               - \frac{1292}{9}\,\zeta(3) 
               - \frac{640}{9}\,\zeta(5)
               + \nf\bigg(
                   -\frac{212}{9} 
                   + \frac{8}{3}\,\zeta(3)
                   \bigg) 
\nonumber\\&\mbox{\hspace{3em}} 
                + \logmuqms\,\bigg(
                   \frac{1430}{3} 
                   - \frac{116}{9}\,\nf
                   \bigg) 
                + \logmuqms^2\,\bigg(
                   108 
                   - \frac{8}{3}\,\nf
                   \bigg) 
               \bigg]
           \Bigg]
\nonumber\\&\mbox{\hspace{1em}} 
 \pm 2\delta_{kM}\delta_{lM}\,
        \Bigg[
           4\,\logmuqms 
                + \api\,\bigg(
               8 
               - 16\,\zeta(3)
               + \frac{16}{3}\,\logmuqms 
               + 8\,\logmuqms^2 
               \bigg)
\nonumber\\&\mbox{\hspace{2em}} 
                + \api^2\,\bigg[
               -\frac{6713}{108} 
               - \frac{464}{3}\,\zeta(3) 
               - \frac{4}{3}\,\zeta(4) 
               + \frac{940}{9}\,\zeta(5)
               + \nf \bigg(
                   \frac{31}{54} 
                   + \frac{8}{9}\,\zeta(3)
                   \bigg) 
\nonumber\\&\mbox{\hspace{3em}} 
                + \logmuqms\,\bigg(
                   \frac{1429}{18} 
                   - \frac{332}{3}\,\zeta(3) 
                   + \nf \left(
                       -3 
                       + \frac{8}{3}\,\zeta(3)
                       \right)
                   \bigg) 
                + \logmuqms^2\,\bigg(
                   \frac{155}{3} 
                   - \frac{14}{9}\,\nf
                   \bigg) 
\nonumber\\&\mbox{\hspace{3em}} 
                + \logmuqms^3\,\bigg(
                   18 
                   - \frac{4}{9}\,\nf
                   \bigg) 
               \bigg]
           \Bigg]
\nonumber\\&\mbox{\hspace{1em}} 
        + (\delta_{kM} + \delta_{lM})
        \Bigg[
           -4 
           - 4\,\logmuqms 
                + \api\,\bigg(
               -\frac{100}{3} 
               + 16\,\zeta(3)
               - \frac{64}{3}\,\logmuqms 
               - 8\,\logmuqms^2 
               \bigg)
\nonumber\\&\mbox{\hspace{2em}} 
                + \api^2\,\bigg[
               -\frac{37123}{108} 
               + \frac{2038}{9}\,\zeta(3) 
               + \frac{4}{3}\,\zeta(4) 
               - \frac{620}{9}\,\zeta(5)
               + \nf \bigg(
                   \frac{605}{54} 
                   - \frac{20}{9}\,\zeta(3)
                   \bigg) 
\nonumber\\&\mbox{\hspace{3em}} 
                + \logmuqms\,\bigg(
                   -\frac{5719}{18} 
                   + \nf \left(
                       \frac{85}{9} 
                       - \frac{8}{3}\,\zeta(3)
                       \right) 
                   + \frac{332}{3}\,\zeta(3)
                   \bigg) 
\nonumber\\&\mbox{\hspace{3em}} 
                + \logmuqms^2\,\bigg(
                   -\frac{317}{3} 
                   + \frac{26}{9}\,\nf
                   \bigg) 
                + \logmuqms^3\,\bigg(
                   -18 
                   + \frac{4}{9}\,\nf
                   \bigg) 
               \bigg]
           \Bigg]
\nonumber\\&\mbox{\hspace{2em}} 
                + \api^2\nh\,\Bigg[
           \bigg(
               \frac{32}{9} 
               + 8\,\logmuqms
               \bigg)
        (\delta_{kM} + \delta_{lM})
           \pm \bigg(
               -\frac{64}{9} 
               - 16\,\logmuqms
               \bigg)\,\delta_{kM}\delta_{lM}
           \Bigg]
\bigg\}\,.\,\,\,\,\,\,\,\,\,\,\,\,
\label{c6val}
\end{align}
\end{subequations}

\subsection{Scalar and pseudo-scalar currents}

Also the results for the scalar and the pseudo-scalar currents (upper
and lower signs, respectively) are taken from \citere{Harlander:diss}:
\begin{subequations}
\begin{align}
&       \hat C^{s/p}_1 =
        \frac{1}{8}\,\api
                + \api^2\bigg[
           \frac{11}{16} 
           + \frac{1}{4}\,\logmuqms
           \bigg] \,,
\label{c1sp}
\\[.5em]
       & \hat C^{s/p}_{2,N} = 
           \api^2\bigg[
               -\frac{5}{6} 
               + \frac{1}{2}\,\logmuqms
               \bigg]\,,
\\[.5em]
       & \hat C^{s/p}_{2,M} = 
           \frac{1}{2} 
                + \api\,\bigg[
               \frac{11}{6} 
                + \logmuqms
                \bigg]
\nonumber\\&\mbox{\hspace{1em}}
                + \api^2\,\bigg[
               \frac{5437}{288} 
               - \frac{17}{4}\,\zeta(3)
               - \frac{79}{144}\,\nf 
                + \logmuqms\,\left(
                   \frac{155}{12} 
                   - \frac{4}{9}\,\nf
                   \right) 
                + \logmuqms^2\,\left(
                   \frac{19}{8} 
                   - \frac{1}{12}\,\nf
                   \right) 
               \bigg]\,,\\
 & \hat C^{s/p}_{2,\text{nd}} = 
                \pm\bigg(1 
                + \api\,\bigg[
               {14\over 3} 
               + 2\,\logmuqms
               \bigg] 
\nonumber\\&\mbox{\hspace{1em}}
                + \api^2\,\bigg[
               {7549\over 144} 
               - {15\over 2}\,\zeta(3)
               - {41\over 24}\,\nf 
                + \logmuqms\,\left(
                   {367\over 12} 
                   - {19\over 18}\,\nf
                   \right) 
                + \logmuqms^2\,\left(
                   {19\over 4} 
                   - {1\over 6}\,\nf
                   \right) 
               \bigg] \bigg)\,,
\\[.5em]
       & \hat C^{s/p}_3 = \frac{3}{16\pi^2}\,\bigg\{
       \delta_{kM}\delta_{lM}\,\Bigg[
           4 
                + \api\,\bigg[
               \frac{32}{3} 
               + 24\,\logmuqms 
               + 16\,\zeta(3)
               \bigg] 
\nonumber\\&\mbox{\hspace{2em}} 
                + \api^2\,\bigg[
               \frac{9955}{36} 
               + \frac{724}{9}\,\zeta(3) 
               - \frac{610}{9}\,\zeta(5)
               + \nf\,\left(
                   -\frac{463}{54} 
                   + \frac{32}{9}\,\zeta(3)
                   \right) 
\nonumber\\&\mbox{\hspace{3em}} 
                + \logmuqms\,\left(
                   \frac{583}{3} 
                   + 140\,\zeta(3)
                   + \nf\,\left(
                       -\frac{46}{9} 
                       - \frac{8}{3}\,\zeta(3)
                       \right) 
                   \right) 
                + \logmuqms^2\,\left(
                   105 
                   - 2\,\nf
                   \right) 
               \bigg]
           \Bigg]
\nonumber\\&\mbox{\hspace{1em}}
     \pm 2\delta_{kM}\delta_{lM}\,\Bigg[
           -4\,\logmuqms 
                + \api\,\bigg[
               -16 
               + 16\,\zeta(3)
               - 24\,\logmuqms 
               - 16\,\logmuqms^2 
               \bigg] 
\nonumber\\&\mbox{\hspace{2em}} 
                + \api^2\,\bigg[
               -\frac{19003}{108} 
               + \frac{574}{3}\,\zeta(3) 
               + \frac{4}{3}\,\zeta(4) 
               + \frac{50}{9}\,\zeta(5)
               + \nf\,\left(
                   \frac{245}{54} 
                   + \frac{16}{9}\,\zeta(3)
                   \right) 
\nonumber\\&\mbox{\hspace{3em}} 
                + \logmuqms\,\left(
                   -\frac{14399}{36} 
                   + \frac{518}{3}\,\zeta(3)
                   + \nf\,\left(
                       \frac{67}{6} 
                       - \frac{8}{3}\,\zeta(3)
                       \right) 
                   \right) 
\nonumber\\&\mbox{\hspace{3em}} 
                + \logmuqms^2\,\left(
                   -222 
                   + \frac{52}{9}\,\nf
                   \right) 
                + \logmuqms^3\,\left(
                   -53 
                   + \frac{10}{9}\,\nf
                   \right) 
               \bigg]
           \Bigg]
\nonumber\\&\mbox{\hspace{1em}}
 + (\delta_{kM} + \delta_{lM})\,\Bigg[
           1 
           - 2\,\logmuqms 
                + \api\,\bigg[
               -6 
               + 8\,\zeta(3)
               - 4\,\logmuqms 
               - 8\,\logmuqms^2 
               \bigg] 
\nonumber\\&\mbox{\hspace{2em}} 
                + \api^2\,\bigg[
               -\frac{13343}{432} 
               + \frac{155}{6}\,\zeta(3) 
               + \frac{2}{3}\,\zeta(4) 
               + \frac{190}{9}\,\zeta(5)
               + \nf\,\left(
                   \frac{205}{216} 
                   + \frac{20}{9}\,\zeta(3)
                   \right) 
\nonumber\\&\mbox{\hspace{3em}} 
                + \logmuqms^2\,\left(
                   -\frac{285}{4} 
                   + \frac{37}{18}\,\nf
                   \right) 
                + \logmuqms^3\,\left(
                   -\frac{53}{2} 
                   + \frac{5}{9}\,\nf
                   \right) 
\nonumber\\&\mbox{\hspace{3em}} 
                + \logmuqms\,\left(
                   -\frac{9017}{72} 
                   + \frac{265}{3}\,\zeta(3)
                   + \nf\,\left(
                       \frac{115}{36} 
                       - \frac{4}{3}\,\zeta(3)
                       \right) 
                   \right) 
               \bigg]
           \Bigg]
\nonumber\\&\mbox{\hspace{1em}}
 + \nh\,\api^2\,\bigg[
           -5 
           + 4\,\zeta(3)
           + 4\,\logmuqms 
                - \logmuqms^2 
           \bigg]
\nonumber\\&\mbox{\hspace{1em}}
  + \nh\,\api^2\,\bigg[
           (\delta_{kM} + \delta_{lM})\,\left(
               -\frac{86}{9} 
               - 8\,\zeta(3)
               + 4\,\logmuqms 
               \right)
\nonumber\\&\mbox{\hspace{3em}}
           \pm \delta_{kM}\delta_{lM}\,\left(
               -\frac{32}{9} 
               - 16\,\zeta(3)
               + 16\,\logmuqms 
               \right)
           \bigg]
           \bigg\}\,.
\label{c6sp}
\end{align}
\end{subequations}

\section{Renormalized mixing matrix}\label{app:zeta22}

For the reader's convenience, we provide the relation between the mixing
matrix $\zeta$ used in this paper and its definition in
\citere{Harlander:2018zpi}, referred to as ``\hkl'' in what
follows. This is easily derived from the relation between the operators
$\calo_{n}$ defined in \eqn{eq:calo} and the $\calo_{n,\mu\nu}$ of
\hkl:
\begin{equation}\label{eq:kj3uo}
  \begin{split}
    \left(
    \begin{matrix}
      \calo_{1}\\
      \calo_{2}
    \end{matrix}
    \right)\delta_{\mu\nu} =
    H_{2\times 2}(a^\bare_s)\left(
    \begin{matrix}
      \calo_{2,\mu\nu}\\ \calo_{4,\mu\nu}
    \end{matrix}
    \right)\,,
  \end{split}
\end{equation}
with $H_{2\times 2}$ defined in \eqn{eq:rdi24}.  The mixing matrix
between regular and flowed operators in \hkl, restricted to the two
operators which are relevant for this paper, is defined through
\begin{equation}\label{eq:0r8ej}
  \begin{split}
    \left(
    \begin{matrix}
      \tcalo_{2,\mu\nu}(t)\\
      \tcalo_{4,\mu\nu}(t)
    \end{matrix}
    \right) = \zeta^\text{\hkl}_{2\times 2}(t)
    \left(
    \begin{matrix}
      \calo_{2,\mu\nu}\\
      \calo_{4,\mu\nu}
    \end{matrix}
    \right)\,,\qquad
    \zeta^\text{\hkl}_{2\times 2}=
    \left(
    \begin{matrix}
      \zeta_{22}^\text{\hkl} &
      \zeta_{24}^\text{\hkl} \\
      \zeta_{42}^\text{\hkl} &
      \zeta_{44}^\text{\hkl}
    \end{matrix}
    \right)\,,
  \end{split}
\end{equation}
where the $\zeta_{ij}^\text{\hkl}$ are the entries of the full
$4\times4$ mixing matrix of \hkl.  Inserting \eqn{eq:kj3uo} and
\begin{equation}\label{eq:znk32}
  \begin{split}
    \left(
    \begin{matrix}
      \tcalo_{1}(t)\\
      \tcalo_{2}(t)
    \end{matrix}
    \right)\delta_{\mu\nu} =
    H_{2\times 2}(\hat{\mu}^{2\ep}\api)\,\chi(t)\,\left(
    \begin{matrix}
      \tcalo_{2,\mu\nu}(t)\\
      \tcalo_{4,\mu\nu}(t)
    \end{matrix}
    \right)\,,\quad\text{with}\quad
    \chi(t) = \left(
    \begin{matrix}
      1 & 0\\
      0 & \zeta_\chi(t)
    \end{matrix}
    \right)\,,\qquad
  \end{split}
\end{equation}
with $\zeta_\chi(t)$ from \eqn{eq:zfchi}, gives the renormalized mixing
matrix used in the current paper in terms of the bare mixing matrix of
\hkl:
\begin{equation}
  \begin{split}
    \zeta_{2\times 2}(t) = H_{2\times 2}(\api\hat{\mu}^{2\ep})\,\chi(t)\,
    \zeta^\text{\hkl}_{2\times 2}(t)\,H_{2\times
      2}^{-1}(\api^\bare)Z_{2\times 2}^{-1}(\api)\,.
    \label{eq:zetahkl}
  \end{split}
\end{equation}
Explicitly, one finds:
\begin{subequations}
  \label{eq:zetaren}
\begin{align}
\zeta_{11}(t) &=    1 
  + \frac{7}{8}\,\api\,\cca 
  + \api^2\,\Bigg\{
      \cca^2\,\bigg[
          \frac{227}{180} 
          - \frac{87}{80}\,\ln2 
          + \frac{27}{40}\,\ln3 
          + \frac{3}{32}\,\lmut
          \bigg] 
\nonumber\\&\hspace*{-2em}
      + \cca\ctr \nf\,\bigg[
          -\frac{1}{18} 
          + \frac{1}{8}\,\lmut
          \bigg] 
      + \ccf\ctr \nf\,\bigg[
          \frac{3}{16} 
          + \frac{1}{4}\,\lmut
          \bigg]
      \Bigg\}\,,\\
      \zeta_{12}(t) &=
\api\,\ccf\,\bigg[
      -\frac{5}{4} 
      - \frac{3}{2}\,\lmut
      \bigg] 
  + \api^2\,\bigg[
      \ccf^2\,\bigg(
          -\frac{17}{32} 
          - \frac{3}{8}\,\lmut
          \bigg) 
\nonumber\\&\hspace*{-2em}
      + \cca\ccf\,\bigg(
          -\frac{431}{60} 
          + \frac{1}{2}\,\zeta(2)
          - \frac{4273}{120}\,\ln2 
          + \frac{2139}{80}\,\ln3 
          + \frac{1}{8}\,\dilogv 
\nonumber\\&\hspace*{-1em}
          - \frac{367}{48}\,\lmut 
          - \frac{11}{8}\,\lmut^2 
          \bigg)
      + \ccf\ctr\nf\,\bigg(
          \frac{15}{8} 
          + \frac{1}{2}\,\zeta(2)
          + \frac{5}{3}\,\lmut 
          + \frac{1}{2}\,\lmut^2 
          \bigg) 
      \bigg]\,,\\
\zeta_{21}(t) &=
\frac{5}{12}\,\api\,\ctr\nf 
  + \api^2\,\bigg[
      \cca\ctr\nf\,\bigg(
          \frac{209}{480} 
          + \frac{9}{20}\,\ln2 
          - \frac{27}{80}\,\ln3
          \bigg) 
\nonumber\\&\hspace*{1em}
      + \ccf\ctr\nf\,\bigg(
          \frac{1}{4} 
          + \frac{1}{2}\,\dilogv 
          - \frac{1}{4}\,\zeta(2)
          + \frac{10}{3}\,\ln2 
          - \frac{21}{8}\,\ln3 
          \bigg)
          \bigg]\,,
  \\
  \zeta_{22}(t) &=
 1 
  + \api\,\ccf\,\bigg[
      \frac{1}{8} 
      - \ln2 
      - \frac{3}{4}\,\ln3
      \bigg]
\nonumber\\&\hspace*{-2em}
  + \api^2\,\Bigg\{
      \cca\ccf\,\bigg[
          -\frac{691}{384} 
          - \frac{55}{8}\,\ln2 
          + \frac{1}{4}\,\ln^22 
          + \frac{63}{16}\,\ln3 
\nonumber\\&\hspace*{0em}
          + \bigg(
              \frac{11}{96} 
              - \frac{11}{12}\,\ln2 
              - \frac{11}{16}\,\ln3
              \bigg)\,\lmut 
          + \frac{3}{4}\,\dilogv 
          - \frac{11}{32}\,\zeta(2)
          \bigg] 
\nonumber\\&\hspace*{0em}
      + \ccf\ctr\nf\,\bigg[
          \frac{333}{160} 
          - \frac{763}{60}\,\ln2 
          + \frac{249}{40}\,\ln3 
          + \bigg(
              -\frac{2}{3} 
              + \frac{1}{3}\,\ln2 
              + \frac{1}{4}\,\ln3
              \bigg)\, \lmut 
\nonumber\\&\hspace*{2em}
          + \frac{1}{4}\,\dilogv 
          - \frac{1}{8}\,\zeta(2)
          \bigg] 
      + \ccf^2\,\bigg[
          \frac{47}{128} 
          + \frac{113}{8}\,\ln2 
          + \frac{1}{4}\,\ln^22 
          - \frac{219}{32}\,\ln3 
\nonumber\\&\hspace*{2em}
          + \frac{51}{16}\,\dilogv 
          + \frac{13}{32}\,\zeta(2)
          \bigg] 
      + \frac{1}{16}\,c_\chi^{(2)}
      \Bigg\}\,,
\end{align}
\end{subequations}
where $c_\chi^{(2)}$ is given in \eqn{eq:n26a3}. The results including
higher orders in $\ep$ are provided in the ancillary file to this paper,
see \app{app:anc}.

\section{Ancillary File}\label{app:anc}

The main results of this paper are provided in computer readable format
(e.g.\ with \texttt{Mathematica}\cite{Mathematica}). The notation is
described in Table\,\ref{tab:anc}. All coefficients are represented by
floating-point numbers in this file. The relative uncertainty for our
numerically evaluated coefficients is estimated to be $10^{-10}$ or
better.

\begin{table}
\begin{center}
\caption[]{\label{tab:anc} Notation in the ancillary file.}
\begin{tabular}{r|cccccc}
  &
  $\zeta_{1}^{(0)}$ &
  $\zeta_{1}^{(2)}$ &
  $\zeta_{13}$ &
  $\zeta_{2}^{(0)}$ &
  $\zeta_{2}^{(2)}$ &
  $\zeta_{23}$\\\hline
  Eq. & 
  \noeqn{eq:zeta01} &
  \noeqn{eq:zeta122a}&
  \noeqn{eq:sq2tea}&
  \noeqn{eq:9s23o}&
  \noeqn{eq:zeta122b}&
  \noeqn{eq:sq2teb}
  \\\hline
  \texttt{code}&
  \texttt{zeta01anc} & 
  \texttt{zeta21anc} & 
  \texttt{zeta13anc} & 
  \texttt{zeta02anc} & 
  \texttt{zeta22anc} & 
  \texttt{zeta23anc}
\end{tabular}\\[2em]
\begin{tabular}{c|ccc}
  &
  $\gamma^\text{f}_{2\times2}$ &
  $\vec{\gamma}^\text{f}_{3}$ &
  $\zeta_{2\times 2}$\\\hline
  Eq. &
  \noeqn{eq:gamma} &
  \noeqn{eq:gammaf3} &
  \noeqn{eq:zetaren} \\\hline
  \texttt{code}&
  \texttt{gamma22anc} &
  \texttt{gamma3anc} &
  \texttt{ZetaMatrix22anc}
\end{tabular}
\end{center}
\end{table}

\end{appendix}

\newcommand{\bibentry}[4]{#1, {\it #2}, #3\ifthenelse{\equal{#4}{}}{}{, }#4.}
\newcommand{\journal}[5]{\href{https://dx.doi.org/#5}{\textit{#1} {\bf #2} (#3) #4}}
\newcommand{\arxiv}[2]{\href{https://arXiv.org/abs/#1}{\texttt{arXiv:#1\,[#2]}}}
\newcommand{\arxivhepph}[1]{\href{https://arXiv.org/abs/hep-ph/#1}{\texttt{hep-ph/#1}}}
\newcommand{\arxivhepth}[1]{\href{https://arXiv.org/abs/hep-th/#1}{\texttt{hep-th/#1}}}
\newcommand{\arxivheplat}[1]{\href{https://arXiv.org/abs/hep-lat/#1}{\texttt{hep-lat/#1}}}
\newcommand{\arxivmathph}[1]{\href{https://arXiv.org/abs/math-ph/#1}{\texttt{math-ph/#1}}}
\newcommand{\arxivmath}[1]{\href{https://arXiv.org/abs/math/#1}{\texttt{math/#1}}}
\IfFileExists{./\jobname_ref.tex}{%

}{}

\end{document}